\documentclass[12pt]{article}
\usepackage{amsmath}
\usepackage{amssymb}
\begin{document}
\begin{center}
{\bf {\large{Constraints, Conserved Charges and Extended BRST Algebra for a 3D Field-Theoretic Example for Hodge Theory}}} 

\vskip 3cm

{\sf  Bhagya. R$^{(a)}$, Harsha Sreekumar$^{(a)}$,  E. Harikumar$^{(a)}$, R. P. Malik$^{(b,c)}$}\\

\vskip 0.1cm

$^{(a)}$ {\it School of Physics, University of Hyderabad,}\\
{\it Central University P. O., Gachibowli, Hyderabad - 500 046, Telangana, India}\\

\vskip 0.1cm

$^{(b)}$ {\it Physics Department, Institute of Science,}\\
{\it Banaras Hindu University, Varanasi - 221 005, Uttar Pradesh, India}\\

\vskip 0.1cm

$^{(c)}$ {\it DST Centre for Interdisciplinary Mathematical Sciences,}\\
{\it Institute of Science, Banaras Hindu University, Varanasi - 221 005, India}\\
{\small {\sf {e-mails: r.bhagya1999@gmail.com; 
harshasreekumark@gmail.com; eharikumar@uohyd.ac.in; rpmalik1995@gmail.com}}}
\end{center}

\vskip 1.5 cm

\noindent
{\bf Abstract:}
We perform the constraint analysis of a three (2 + 1)-dimensional (3D) field-theoretic example for Hodge theory (i) at the {\it classical}
level within the ambit of Lagrangian formulation, and
(ii) at the {\it quantum} level within the framework of Becchi-Rouet-Stora-Tyutin (BRST) formalism. We derive the conserved charges 
corresponding to the {\it six} continuous symmetries of our present theory. These six continuous summery transformations are the
nilpotent (anti-)BRST and (anti-)co-BRST symmetries, a unique bosonic symmetry and the ghost-scale symmetry. It turns out that
the {\it Noether} conserved (anti-)BRST charges are found to be non-nilpotent even though they are derived from the off-shell nilpotent versions of the
continuous and infinitesimal (anti-)BRST symmetry transformations. We obtain the nilpotent versions of the (anti-)BRST charges 
from the non-nilpotent {\it Noether} (anti-)BRST charges and discuss the physicality criteria w.r.t. the {\it latter}  to demonstrate that
the operator forms of the first-class constraints (of the {\it classical} gauge theory) annihilate the physical states at
the quantum level. This observation is consistent with Dirac's quantization conditions for the systems that are endowed with the constraints.
We lay emphasis on the  existence of a {\it single} (anti-)BRST invariant Curci-Ferrari (CF) type restriction  in our theory and derive it from various 
theoretical angles. \\





\newpage

\section{Introduction}

Over the years, the key concepts of pure mathematics  and their powerful impact on the modern developments in the realm 
of theoretical physics have been intertwined together in a meaningful manner. In particular, the modern upsurge of interest 
in the domain of (super)string theories has brought together the top-class mathematicians and theoretical physicists on
a single platform (see, e.g. [1-5] and references therein) where the knowledge and understandings of {\it both} types
of researchers have  been enriched. During the past few years, we have devoted significant amount of time
on the study of field-theoretic as well as the toy models of Hodge theory (see, e.g. [6-10] and references therein) within
the framework of Becchi-Rouet-Stora-Tyutin (BRST) formalism [11-14]. In this domain of study, there has been convergence 
and confluence of ideas
from the mathematics of the de Rham cohomological operators of differential geometry (see, e.g. [15-18]) and the physical
aspects of the BRST formalism (where the discrete and continuous symmetries and corresponding conserved charges have been given utmost importance). To be precise,
we have been able to establish that the {\it massless} and the St${\ddot u}$ckelberg-modified {\it massive} Abelian
$p$-form (with $p = 1, 2, 3 $)  gauge theories are the tractable field-theoretic examples for Hodge theory in the
$D = 2 p$ (i.e. $D = 2, 4, 6 $) dimensions of spacetime where there is a two-to-one mapping between the 
continuous symmetries (and corresponding conserved charges) and the cohomological operators of the differential geometry 
at the {\it algebraic} level.

In a very recent work [19], we have been able to show that a combination of free Abelian 1-form and 2-form gauge theories
provide a field-theoretic example for Hodge theory in the {\it odd} (i.e. $D = 3 $) dimensions of spacetime where the continuous symmetries 
of {\it this} three (2 + 1)-dimensional (3D) field-theoretic system 
have provided the physical realizations of the de Rham cohomological
operators of differential geometry at the {\it algebraic} level. However, in [19], we have {\it not} been able to
incorporate the Noether conserved charges (corresponding to the above continuous symmetries) and the algebra satisfied by them.
One of the central purposes of our present endeavor is (i) to compute all the Noether conserved charges corresponding
to the {\it six} continuous symmetry transformations that exist for the coupled (but equivalent) (anti-)BRST
invariant Lagrangian densities [cf. Eqs. (36),(16) below] of our present theory, (ii) to derive the extended BRST algebra among
the conserved (but {\it appropriate}) forms of the charges, and (iii) to establish that there is a two-to-one mapping between
the appropriate forms of the conserved charges and the cohomological operators of  differential geometry. The study of the
field-theoretic models of Hodge theory is {\it physically} important and useful because we have been able to show that
(i) the two (1 + 1)-dimensional (2D) (non-)Abelian gauge theories (without any interactions with the matter fields) are the
examples of a {\it new} type [20] of topological field theory (TFT) which captures a few aspects of the Witten-type TFTs [21] and
some salient features of the Schwarz-type TFTs [22], and (ii) the 4D free Abelian 2-form and  6D free Abelian 3-form theories
provide a set of models for quasi-TFTs (see, e.g. [23,8] for details) where there is existence of the topological invariants
(with their proper recursion relationships) but the Lagrangian densities of the above theories are neither like [21] nor like [22].

Against the backdrop of the discussions in the above paragraphs where a few results have been pointed out,  we assimilate here
the key results of our present endeavor together for the readers' convenience. For the first-time, we have been able
provide an odd-dimensional (i.e. $ D = 3$) field-theoretic example for Hodge theory\footnote{We have been
able to prove that any arbitrary massless and St${\ddot u}$ckelberg-modified massive Abelian $p$-form ($p =1, 2, 3... $) 
gauge theories in the $D = 2 p$ (i.e. $D = 2, 4, 6... $) dimensions of spacetime are the field-theoretic examples for 
Hodge theory within the framework of BRST formalism. However, it is clear that {\it such} theories are defined 
{\it only} in the even  (i.e. $D = 2 p \equiv 2, 4, 6...$)  dimensions of spacetime.}    
where the continuous symmetries and corresponding conserved charges have been discussed {\it together} 
within the framework of Becchi-Rouet-Stora-Tyutin (BRST) formalism. We have demonstrated the existence of
the first-class constraints for our {\it classical} 3D combined system of the Abelian 1-form and 2-form gauge 
theories and established {\it their} connection with the Noether conserved charge that turns out to be the generator for 
the {\it classical} gauge symmetry transformations that exist in the theory (due to the presence of the above first-class constraints).
We have generalized (i) the classical 3D Lagrangian density  to its counterparts (anti-)BRST invariant
coupled (but equivalent) Lagrangian densities [cf. Eqs (36),(16) below] at the quantum level, and (ii) the
local, infinitesimal and continuous {\it classical} gauge symmetry transformations to their
{\it quantum} counterparts (i.e. infinitesimal, continuous and off-shell nilpotent versions of the (anti-)BRST
symmetry transformations [cf. Eqs. (37),(17) below]). The above (anti-)BRST invariant Lagrangian densities
respect {\it six} continuous symmetry transformations [19]. We have computed corresponding {\it appropriate} conserved charges
and derived the extended BRST algebra that is reminiscent of the Hodge algebra satisfied by the de Rham cohomological
operators [15-18]. The physicality criteria w.r.t. the
appropriate conserved and nilpotent (anti-)BRST charges have been discussed thoroughly in our present investigation.
Finally, we have been able to obtain a two-to-one mapping between the {\it appropriate} 
conserved charges and the cohomological operators at the {\it algebraic} level.

Our present investigation is interesting, important and essential on the following counts. First of all, even though we have
been able to demonstrate an odd dimensional (i.e. $ D = 3$) field-theoretic model to be an example for Hodge theory in our earlier work [19],
we have discussed {\it only} a set of {\it six} continuous symmetry transformations. However, we have {\it not} derived
the corresponding Noether conserved charges in [19]. One of the key motivations of our present endeavor is (i) to compute 
all the Noether conserved charges corresponding to all the  above six continuous symmetry transformations, and (ii) to discuss
the characteristic features that are associated with these charges. Second, we have provided the physical realization(s) of the
de Rham cohomological operators of differential geometry [15-18] in terms of the symmetry operators in our earlier work [19]. In
our present endeavor we derive the extended BRST algebra among the {\it appropriate} versions of the conserved  charges
and provide the physical realization(s) of the above cohomological operators in the language of these charges. Finally, we discuss
the physicality criteria w.r.t. the off-shell nilpotent  versions of the (anti-)BRST charges  and establish their {\it consistency}
with the Dirac quantization conditions for the physical systems that are endowed with the constraints. In other words, we explicitly
show that the operator forms of the first-class constraints of our 
{\it classical} 3D gauge theory  annihilate the {\it true} physical states 
(which are present in the  {\it total} quantum Hilbert space of states) at the quantum level.

The theoretical materials of our present endeavor are organized in the following order. We perform, in Sec. 2, 
the constraint analysis for the combined system of the D-dimensional free Abelian 1-form and 2-form theories.
In our Sec. 3, we generalize the 3D {\it classical} version of the Lagrangian density (of our previous section) to
the {\it quantum} level within the framework of BRST formalism and discuss the off-shell nilpotent BRST symmetries, derive the 
corresponding Noether conserved charge and discuss the nilpotency property of the {\it latter}. Our Sec. 4 is devoted
to the derivation of the off-shell nilpotent anti-BRST symmetries and corresponding 
Noether anti-BRST charge. The central subject matter of our Sec. 5 is the derivation of the
(anti-)co-BRST symmetries and corresponding Noether conserved charges. In Sec. 6, we derive the bosonic symmetry
transformations that are the appropriate anticommutators of the above off-shell nilpotent symmetries, show the
{\it uniqueness} of  these transformations and derive the corresponding conserved charge.  Our Sec. 7 contains
the ghost-scale symmetry transformations and corresponding Noether conserved charge. In Sec. 8, we devote time on the 
derivation of the extended BRST algebra that is obeyed by the above 
{\it appropriate} versions of the conserved charges. Finally, in Sec. 9, we summarize our
key results and comment on the future perspective of our present endeavor.

In our Appendix A, we derive the CF-type restriction from the direct equality of the (anti-)BRST
invariant Lagrangian  densities [cf. Eqs. (36),(16)]. We deal with the absolute
anticommutativity property of the nilpotent and conserved (anti-)BRST charges in our Appendix B. 
Our Appendix C is devoted to the proof of the anticommutativity property 
of the off-shell nilpotent and conserved (anti-)co-BRST [i.e. (anti-)dual BRST] charges. In our Appendix D, 
we derive the expression for the bosonic charge separately and independently from two anticommutators which are defined in terms of the 
off-shell nilpotent versions of the conserved (anti-)BRST charges and (anti-)co-BRST charges. \\

{\it Conventions and Notations:} Our 3D background flat Minkowskian spacetime manifold is endowed with the metric tensor
$\eta_{\mu\nu} = $ diag ($+1, -1, -1$) so that the dot product between two non-null vectors $P_\mu$ and $Q_\mu$ is
defined as: $P \cdot Q = \eta_{\mu\nu}\, P^\mu\, Q^\nu \equiv P_0 Q_0 - P_i Q_i$ where the Greek indices $\mu, \nu, \sigma...= 0, 1, 2$ stand for the time
and space directions and the Latin indices $i, j, k...= 1, 2 $ correspond to the space directions {\it only}. 
We adopt the convention of the {\it left} derivative w.r.t. the fermionic fields of our theory in
all the computations where such derivatives are required. We have also adopted the convention of derivative w.r.t. the
antisymmetric tensor gauge field as: $(\partial B_{\mu\nu}/ \partial B_{\sigma\eta}) = \frac{1}{2!}\, 
(\delta^\sigma_\mu \delta^\eta_\nu - \delta^\sigma_\nu \delta^\eta_\mu)$.
The notations $s_{(a)b}$ and $s_{(a)d}$ have been chosen for the
off-shell nilpotent (i.e. $s_{(a)b}^2 = 0, \, s_{(a)d}^2 = 0 $) versions of the (anti-)BRST and (anti-)dual BRST [i.e. (anti-)co-BRST]
symmetry transformations, respectively. Similarly, we adopt the symbols $Q_{(a)b}$  and $Q_{(a)d}$ to denote
the {\it Noether} conserved (anti-)BRST and (anti-)co-BRST charges, respectively. The overdot notation on a generic field
(i.e. $\dot \Phi $) has been chosen to express the partial time derivative (i.e. $\partial \Phi/\partial t$)  on it.
Throughout the whole body of our text, we have taken into account the natural units: $\hbar = c = 1$ to define the 
(anti)commutators as well as the time derivative.


\section{Preliminary: Constraints of Our Free Theory at the Classical Level in the Lagrangian Formulation }

We begin with the Lagrangian density [${\cal L}_{(0)}$] for the
combined system of the D-dimensional free Abelian 1-form and 2-form gauge field theories as 
\begin{eqnarray}
{\cal L}_{(0)} = -\; \frac{1}{4}\; F^{\mu\nu}\, F_{\mu\nu} + \frac{1}{12}\, H^{\mu\nu\sigma}\, H_{\mu\nu\sigma},
\end{eqnarray}
where the field-strength tensors: $F_{\mu\nu} = \partial_\mu \, A_\nu -\partial_\nu \, A_\mu$  and
$H_{\mu\nu\sigma} = \partial_\mu \, B_{\nu\sigma} + \partial_\nu \, B_{\sigma\mu} + \partial_\sigma \, B_{\mu\nu}$
are derived from the 2-form $ F^{(2)} = d\, A^{(1)} = \frac{1}{2!}\, F_{\mu\nu}\, (dx^\mu \wedge dx^\nu)$  and 3-form
$ H^{(3)} = d\, B^{(2)} = \frac{1}{3!}\, H_{\mu\nu\sigma}\, (dx^\mu \wedge dx^\nu \wedge dx^\sigma)$
where $d = \partial_\mu \;dx^\mu$ [with $d^2 = \frac{1}{2!}\, (\partial_\mu \partial_\nu - \partial_\nu \partial_\mu)\, (dx^\mu \wedge dx^\nu) = 0 $]
is the exterior derivative of differential geometry (see, e.g. [15-18]) and the 1-form $A^{(1)} = A_\mu\, dx^\mu $ 
and 2-form $ B^{(2)} =\frac{1}{2!}\, B_{\mu\nu}\, (dx^\mu \wedge dx^\nu)$  define the vector gauge field $A_\mu$ and
antisymmetric (i.e. $B_{\mu\nu} = -\, B_{\nu\mu} $) tensor gauge field $B_{\mu\nu}$, respectively. Here, as is obvious,
the Greek indices: $\mu, \nu, \sigma...= 0, 1, 2...D-1$. The above Lagrangian density (1) is
{\it singular} w.r.t. the {\it basic} gauge fields of our theory. As a consequence, there are
constraints on the theory which  can be studied, first of all, through the definitions of the canonical conjugate
momenta $\Pi^\mu_{(A)}$ and $\Pi^{\mu\nu}_{(B)}$ w.r.t. the gauge fields $A_\mu$ and $B_{\mu\nu}$, respectively. These conjugate
momenta, for our present combined system of the Abelian 1-form and 2-form gauge theories, are:
\begin{eqnarray} 
&&\Pi^\mu_{(A)}   = \frac{\partial\, {\cal L}_{(0)}}{\partial\, (\partial_0\, A_\mu)} = -\, F^{0\mu}
\; \quad  \Longrightarrow \quad \;\Pi^0_{(A)}  =  -\, F^{00} \approx 0,   \nonumber\\ 
&&\Pi_{(B)}^{\mu\nu}  = \frac {\partial {\cal L}_{(0)}}{\partial (\partial_0 B_{\mu\nu})}
 = \frac {1}{2}\, H^{0\mu\nu} \quad  \Longrightarrow \quad
\Pi^{0i}_{(B)}  = \frac {1}{2}\, H^{00i} \approx 0.
\end{eqnarray}
A close look at the above equation demonstrates that we have two primary constraints $\Pi^0_{(A)} \approx 0 $ and $\Pi^{0i}_{(B)} \approx 0$
on our theory where we have used the symbol $\approx 0 $ to denote Dirac's notation for the constraints to be weakly zero. As a consequence, we
are allowed to take a first-order time derivative on the above primary constraints (PCs) and set it
equal to zero. Physically, this is the requirement for the time-evolution invariance of the PCs which leads to the 
precise determination of the secondary constraints (SCs) 
on the theory.  The well-known Hamiltonian formulation is the most suitable approach to deal with the constraint analysis. However, for
the simple systems (like our present 3D model) the Lagrangian formulation is good enough (see, e.g. [24]). To obtain the secondary constraints
on our theory, we have to focus on the time-evolution invariance of the PCs which can be derived from the following Euler-Lagrange (EL)
equations of motion (EoM)
\begin{eqnarray}
\partial_\mu\, F^{\mu\nu} = 0, \qquad \quad \partial_\mu\, H^{\mu\nu\sigma} = 0, 
\end{eqnarray}
which emerge out from the starting Lagrangian density (1). To accomplish the above goals
of the determination of the SCs, we take into account the choices $ \nu = 0$ in
the first-entry and $\nu = 0, \, \sigma = i$ in the second-entry of the above EL-EoMs which lead to the following
\begin{eqnarray}
&&\partial_\mu\, F^{\mu0} = 0 \quad \Longrightarrow \quad \partial_0 F^{00} + \partial_i F^{i0} = 0,  \nonumber\\
&& \partial_\mu\, H^{\mu  0 i} = 0 \quad \Longrightarrow \quad \partial_0 H^{00i} + \partial_j H^{j0i} = 0,
\end{eqnarray}
where we have taken into account (as per Dirac's prescription) the fact that the PCs are weakly zero and the first-order time
derivative on them is allowed. Substitutions of the explicit expressions for the components of the momenta [cf. Eq. (2)]
of our free 3D theory  into
the above equation leads to the following relationships:
\begin{eqnarray}
{\displaystyle \frac{\partial \Pi^0_{(A)}}{\partial t}} = \partial_i \, \Pi^i_{(A)} \equiv \partial_i E_i \approx 0,   \qquad
{\displaystyle \frac{\partial \Pi^{0i}_{(B)}}{\partial t}} = \partial_j \, \Pi^{ji}_{(B)} \approx 0. 
\end{eqnarray}
A close look at the above equation demonstrates that we have already obtained the SCs as: $\partial_i \, \Pi^i_{(A)} \approx 0,\,
 \partial_i \, \Pi^{ij}_{(B)} \approx 0$. In the above, we have taken into account the symbols: $ \Pi^i_{(A)} = -\, F^{0i} \equiv E_i, \;
\Pi^{ij}_{(B)} = \frac{1}{2}\, H^{0ij} $ which are nothing but the space components of the covariant expressions for the
canonical conjugate momenta [cf. Eq. (2)] w.r.t. the basic gauge fields $A_\mu$ and $B_{\mu\nu}$, respectively. In particular, we
note that $E_i = (E_1, E_2)$ is nothing but the electric field for the 3D Abelian 1-form theory (with two existing components). At this juncture,
we point out that a couple of primary constraints (i.e. $\Pi^0_{(A)} \approx 0, \Pi^{0i}_{(B)} \approx 0$)
and another couple of secondary constraints (i.e. $\partial_i \, \Pi^i_{(A)} \approx 0,\, \partial_i \, \Pi^{ij}_{(B)} \approx 0$) of our theory 
are {\it all} expressed in terms of the components of the canonical conjugate momenta [cf. Eq. (2)]. As a consequence, all of them commute among themselves
which establishes their categorization as the first-class constraints in the well-known terminology of Dirac's prescription for the 
classification scheme  of constraints (see, e.g.[25-29] for details).

Against the backdrop of the above paragraph, it is pertinent to point out that the first-class constraints {\it always} generate the infinitesimal, local
and continuous gauge symmetry transformations. To corroborate this statement, let us, first of all, write down the well-known infinitesimal,
local and continuous  gauge symmetry
transformations ($\delta_g $) for the combined system of the free Abelian 1-form and 2-form theories {\it together} as 
\begin{eqnarray}
\delta_g\, A_\mu = \partial_\mu\, \Lambda,  \qquad \delta_g \, B_{\mu\nu} = -\, \big (\partial_\mu \Lambda_\nu - \partial_\nu \Lambda_\mu \big ),
\qquad \delta_g {\cal L}_{(0)} = 0, 
\end{eqnarray}
where the Lorentz scalar $\Lambda (x) $ and Lorentz vector  $\Lambda_\mu (x) $ are the local infinitesimal gauge 
symmetry transformations parameters. According to
Noether's theorem, the existence of the infinitesimal continuous symmetry transformations for a theory (i.e. $\delta_g {\cal L}_{(0)} = 0 $)
always implies the existence of the Noether conserved current and corresponding conserved charge. The {\it former}  for our theory
(i.e. $J^\mu_{(G)}$) is as follows:
\begin{eqnarray}
J^\mu_{(G)} = - \,F^{\mu\nu}\, \partial_\nu \Lambda -  \frac{1}{2}\; H^{\mu\nu\sigma} \, (\partial_\nu \Lambda_\sigma - \partial_\sigma \Lambda_\nu).
\end{eqnarray}
The conservation law (i.e. $\partial_\mu J^\mu_{(G)} = 0$) for the above current is straightforward provided we use
the EL-EoMs (3) that have been derived from the Lagrangian density (1). The expression for the Noether conserved charge $Q_{(G)}$ (that emerges out
from the above conserved current for our D-dimensional free theory) is as follows:
\begin{eqnarray}
Q_{(G)} = \int d^{D-1} x \, J^0_{(G)} \equiv \int d^{D-1} x \, \Big [ - F^{0\nu} \, \partial_\nu \Lambda \,-\,
\frac{1}{2}\, H^{0\nu\sigma} \, \big ( \partial_\nu\Lambda_\sigma -  \partial_\sigma \Lambda_\nu \big ) \Big].
\end{eqnarray}
Keeping in our mind the fact that our theory is endowed with the primary constraints (i.e.  $\Pi^0_{(A)} = - \, F^{00} \approx 0, 
\Pi^{0i}_{(B)} = \frac{1}{2} \, H^{00i} \approx 0$), we have to expand the r.h.s. of the above equation in such a manner that
these constraints are {\it not strongly} equal to zero. Thus, from (8), we obtain the following expression for $Q_{(G)}$ in terms of
the components of momenta, namely;
\begin{eqnarray}
Q_{(G)} = \int d^{D-1} x \, \Big [ \Pi^0_{(A)} \, \partial_0 \Lambda + \Pi^i_{(A)} \, \partial_i \Lambda - \Pi^{0i}_{(B)} 
\, \big ( \partial_0\Lambda_i -  \partial_i \Lambda_0 \big )  - \Pi^{ij}_{(B)} 
\, \big ( \partial_i\Lambda_j -  \partial_j \Lambda_i \big )\Big],
\end{eqnarray}
where the precise expressions for the components of the momenta have been taken from (2). It is straightforward now to verify that
the application of the following equal-time non-zero canonical commutation relations of our theory, namely;
\begin{eqnarray*}
&&{\left[A_{0}(\vec{x}, t), \Pi_{(A)}^{0}(\vec{y}, t)\right] }  = i\, \delta^{(D-1)}(\vec{x}-\vec{y}),\nonumber\\
&&{\left[A_{i}(\vec{x}, t), \Pi_{(A)}^{j}(\vec{y}, t)\right] }  = i \,\delta_{i}^{j} \,\delta^{(D-1)}(\vec{x}-\vec{y}),\nonumber\\
\end{eqnarray*}
\begin{eqnarray}
&&{\left[B_{0 i}(\vec{x}, t), \Pi_{(B)}^{0 j}(\vec{y}, t)\right] }  = i \, \delta_{i}^{j} \, \delta^{(D-1)}(\vec{x}-\vec{y}), \nonumber\\
&&{\left[B_{i j}(\vec{x}, t), \Pi_{(B)}^{k l}(\vec{y}, t)\right] }  =\frac{i}{2!}\,
\left(\delta_{i}^{k} \delta_{j}^{l}-\delta_{i}^{l} \delta_{j}^{k}\right) \,
\delta^{(D-1)}(\vec{x}-\vec{y}),
\end{eqnarray}
lead to the derivation of the infinitesimal, local and continuous gauge symmetry transformations (6) for the gauge fields. To corroborate
this statement, we have to exploit the theoretical strength of the following standard relationship between the continuous symmetry
transformations $\delta_g$ and the generator $Q_{(G)}$, namely; 
\begin{eqnarray}
\delta_g \, \Phi (\vec{x}, t) =  {\left[ \Phi(\vec{x}, t), Q_{(G)}\right] },
\end{eqnarray}
where $\Phi (\vec{x}, t) $ is the generic field of our gauge theory.

We end this section with the following crucial remarks. First of all, we can recast the above expression for the Noether conserved charge
 $Q_{(G)}$ [cf. Eq. (9)] in terms of the first-class constraints (i.e. $\Pi^0_{(A)} \approx 0, \Pi^{0i}_{(B)} \approx 0, 
\partial_i \, \Pi^i_{(A)} \approx 0,\, \partial_i \, \Pi^{ij}_{(B)} \approx 0$) by performing the partial integration and dropping
the total space derivative terms due to Gauss's divergence theorem. The final form $G$ of the above Noether conserved charge
$Q_{(G)}$, in terms of the first-class constraints, can be explicitly expressed as follows
\begin{eqnarray}
Q_{(G)} \to G &=& \int d^{D-1} x \, \Big [ \Pi^0_{(A)} \, \partial_0 \Lambda  - (\partial_i \Pi^i_{(A)}) \,  \Lambda - \Pi^{0i}_{(B)} 
\, \partial_0\Lambda_i - \Pi^{i0}_{(B)} \,\partial_i \Lambda_0  \nonumber\\
&+& (\partial_i\Pi^{ij}_{(B)}) \,\Lambda_j + (\partial_j \Pi^{ji}_{(B)}) \,\Lambda_i \Big],
\end{eqnarray}
which matches with the standard relationship between the generator $G$ for the gauge symmetry transformation and the first-class
constraints that has been written down in a very nice paper (see, e.g. [30,31] for details). Second, as far as our present endeavor
on the 3D field-theoretic model is concerned, it is interesting to point out that the kinetic term for the Abelian 2-form field
reduces to the following simple form, namely;
\begin{eqnarray}
\frac{1}{12}\, H^{\mu\nu\lambda}\, H_{\mu\nu\lambda} = \frac{1}{2}\, H^{012}\, H_{012} = \frac{1}{2}\, (H_{012})^2.
\end{eqnarray}
In other words, the totally antisymmetric field-strength tensor 
$H_{\mu\nu\sigma}$ for the Abelian 2-form field has only a {\it single} existing independent component $H_{012}$ which can be written in
its covariant form as: $H_{012} = \frac{1}{2!}\, \varepsilon^{\mu\nu\sigma}\, \partial_\mu\, B_{\nu\sigma} \equiv 
\frac{1}{3!}\, \varepsilon^{\mu\nu\sigma}\, H_{\mu\nu\sigma} $. Hence, the correct form of the Lagrangian density (1)
for our 3D field-theoretic model can be expressed as:
\begin{eqnarray}
{\cal L}^{(3D)}_{(0)} = -\; \frac{1}{4}\; F^{\mu\nu}\, F_{\mu\nu} 
+ \frac{1}{2}\, \Big (\frac{1}{2}\,\varepsilon^{\mu\nu\sigma}\, \partial _\mu B_{\nu\sigma} \Big )^2.
\end{eqnarray}
From the above Lagrangian density, we can define the antisymmetric (i.e. $\Pi_{(B)}^{\mu\nu} = -\, \Pi_{(B)}^{\nu\mu} $)
conjugate momenta w.r.t. the antisymmetric (i.e. $B_{\mu\nu} = -\, B_{\nu\mu}$) 2-form field $B_{\mu\nu}$ as:
\begin{eqnarray}
\Pi_{(B)}^{\mu\nu}  = \frac {\partial {\cal L}^{(3D)}_{(0)}}{\partial (\partial_0 B_{\mu\nu})}
 = \frac {1}{2}\, \varepsilon^{0\mu\nu}\, H_{012} \quad  \Longrightarrow \quad
\Pi^{0i}_{(B)}  = \frac {1}{2}\, \varepsilon^{00i} \,  H_{012}  \approx 0.
\end{eqnarray}
Thus, we have the primary constraint for our 3D theory as: $\Pi^{0i}_{(B)}  = \frac {1}{2}\, \varepsilon^{00i} \,  H_{012}  \approx 0$. 
The secondary constraint, in the above language, would be: $\partial_i \Pi^{ij}_{(B)} \approx 0 $ where the antisymmetric
space components of the conjugate momenta are: $ \Pi^{ij}_{(B)} = \frac {1}{2}\, \varepsilon^{0ij} \,  H_{012}$. Finally, we would like
to mention that, according to Dirac's prescription for the quantization conditions on the 
physical systems with constraints, it is an essential
requirement that the operator forms of the constraints (existing at the {\it classical} level) 
must annihilate the {\it physical} states (exiting in the {\it total} 
quantum Hilbert space of states) at the {\it quantum} level. We shall see that the physicality criteria w.r.t. the 
{\it nilpotent} versions of the (anti-)BRST
charges would lead to the validity of the above conditions (cf. Secs 3 and 4 below for details).

\section{BRST Invariant Lagrangian Density at the Quantum Level: Off-Shell Nilpotent BRST Symmetries}

Our present section is divided into two parts. In Subsec. 3.1, we derive the standard Noether BRST current and corresponding charge from
the off-shell nilpotent BRST symmetry transformations and show that the Noether conserved BRST charge $Q_b$ is non-nilpotent. Starting from
this non-nilpotent (i.e. $Q_b^2 \neq 0 $) version of the {\it Noether} BRST charge $Q_b$, we derive  
the off-shell nilpotent (i.e. $Q_B^2 = 0 $) version of the BRST charge $Q_B$ in Subsec. 3.2.

\subsection{BRST Symmetries and Conserved Noether Charge}

We generalize the classical 3D Lagrangian density (14) to its counterpart BRST-invariant Lagrangian density 
${\cal L}_B$  at the quantum level that incorporates into itself 
the gauge-fixing and Faddeev-Popov (FP) ghost terms as (see, e.g. [19] for details)
\begin{eqnarray}
{\cal L}_{B} &=& -\frac{1}{4} F^{\mu \nu} F_{\mu \nu} +  \frac{B^{2}}{2}   -  B \, (\partial \cdot A)
- \partial_{\mu} \bar{C} \, \partial^{\mu} C  
+ \mathcal{B} \left(\frac{1}{2} \varepsilon^{\mu \nu \sigma} \partial_{\mu} B_{\nu \sigma}\right)  
-\frac{\mathcal{B}^{2}}{2}\nonumber\\
&+& B^{\mu} \left(\partial^{\nu} B_{\nu \mu}-\partial_{\mu} \phi \right)  - \frac{B^{\mu} B_{\mu}}{2}   
+ \partial_{\mu} \,  \bar{\beta} \,\partial^{\mu} \beta +\left(\partial_{\mu} \bar{C}_{\nu} 
-\partial_{\nu} \bar{C}_{\mu}\right)\left(\partial^{\mu} C^{\nu}\right) \nonumber\\ 
&+& \left (\partial \cdot \bar{C}+\rho \right)\,  \lambda + \left (\partial \cdot C - \lambda \right )\, \rho,
\end{eqnarray}
where the first {\it four} terms (i.e. $ -\frac{1}{4} F^{\mu \nu} F_{\mu \nu} + \frac{1}{2}\, B^2  -  B \,(\partial \cdot A) 
- \partial_{\mu} \bar{C} \, \partial^{\mu} C$) belong to the (anti-)BRST invariant Lagrangian density for the free
Abelian 1-form  gauge theory which contains the corresponding gauge-fixing term and FP-ghost term. 
The {\it rest} of the terms of the above Lagrangian density (16) are for the BRST-invariant
Lagrangian density in the case of a 3D free Abelian 2-form theory
(see, e.g. [32,33]). In the BRST invariant Lagrangian density (16), we have the Lorentz vector 
fermionic (i.e. $C_\mu^2 = \bar C_\mu^2 = 0, \, C_\mu\, C_\nu + C_\nu \, C_\mu = 0, \, 
C_\mu \bar C_\nu + \bar C_\nu \, C_\mu = 0, $ etc.) (anti-)ghost fields ($\bar C_\mu)C_\mu$
with the ghost numbers $(-1)+1$, respectively. These (anti-)ghost fields
are the generalizations of the gauge symmetry parameter $\Lambda_\mu$ [cf. Eq. (6)]. On the other hand, the (anti-)ghost fields ($\bar\beta)\,\beta $
are the ghost-for-ghost fields which are bosonic (i.e. $\beta^2 \neq 0, \bar\beta^2 \neq 0$) in nature
and they carry the ghost numbers $(- 2)+2$, respectively. The fermionic (i.e. $C^2 = 0, \, \bar C^2 = 0, \, 
C\, \bar C + \bar C \, C = 0$) (anti-)ghost fields ($\bar C)\, C$  are endowed with the ghost numbers $(-1)+1$,  respectively, which
correspond to the generalizations of the classical gauge symmetry transformation parameter $\Lambda$ [cf. Eq. (6)] to the quantum level. 
These {\it latter} (anti-)ghost fields correspond to the BRST-invariant Abelian 1-form gauge theory (discussed within the framework of BRST formalism). 
The auxiliary (anti-)ghost fields $(\rho)\lambda$  of our system
also carry the ghost numbers $(- 1)+1$, respectively, because we note that 
$\rho = -\, (1/2) \, (\partial \cdot \bar C)$  and $\lambda =  (1/2) \, (\partial \cdot  C)$. 
These (anti-)ghost fields are required to maintain the sacrosanct  property of unitarity  in our BRST-invariant theory
which is valid at any arbitrary order of perturbative  computations for {\it all} 
the physical processes that are allowed by our BRST-quantized theory.

The above {\it quantum} version of the Lagrangian density (16) respects the following infinitesimal, continuous and off-shell nilpotent 
($s_b^2 = 0$) BRST transformations ($s_b$)
\begin{eqnarray}
&& s_{b} B_{\mu \nu}=-\left(\partial_{\mu} C_{\nu} - \partial_{\nu} C_{\mu}\right), \qquad s_{b} C_{\mu}= -\,\partial_{\mu} \beta, \qquad
s_{b} \bar{C}_{\mu}= - \,B_{\mu}, \qquad s_{b} \bar{C}= B,
\nonumber\\
&& s_{b} A_{\mu}= \partial_{\mu} C, \quad  s_{b} \bar{\beta} = - \,\rho, \quad s_{b} \phi = + \,\lambda, \quad 
s_{b}\left[ \rho, \lambda, C, \beta,  B, \mathcal{B}, B_{\mu}, F_{\mu \nu}, H_{\mu \nu \lambda}\right]=0,
\end{eqnarray}
because we observe  that  $\mathcal{L}_{B} $ [cf. Eq. (16)] transforms to a {\it total} spacetime derivative: 
\begin{eqnarray}
s_{b} \mathcal{L}_{B} = - \,\partial_{\mu}\Big[\left(\partial^{\mu} C^{\nu}-\partial^{\nu} C^{\mu}\right) B_{\nu}
+ \lambda \,B^{\mu} 
+ \rho\, \partial^{\mu} \beta + B\,\partial^{\mu} C\Big]. 
\end{eqnarray}
As a consequence of the Gauss divergence theorem, 
the action integral $S = \int d^3 x \, {\cal L}_B$  [corresponding to the Lagrangian density (16)] can be expressed as the surface term
and all the fields will go to infinity. However, 
since all the physical fields vanish off  
 as:  $x \rightarrow \pm \infty$, we find that the action integral
remains invariant (i.e. $s_b\, S = 0$).
  The observation in (18) implies that there is an infinitesimal,  
continuous and off-shell nilpotent {\it symmetry} invariance in the theory which
 leads to the derivation of the following Noether current: 
\begin{eqnarray}
J^\mu_b &=& \left(\partial^{\mu} \bar C^{\nu} - \partial^{\nu} \bar C^{\mu}\right)\, \partial_\nu \beta - B\, \partial^\mu C - F^{\mu\nu}\, \partial_\nu C 
- \lambda \, B^\mu - \rho \,\partial^\mu \beta \nonumber\\
&-& \varepsilon^{\mu\nu\sigma} \, {\cal B} \, \partial_\nu C_\sigma - \left(\partial^{\mu} C^{\nu} - \partial^{\nu} C^{\mu}\right)\, B_\nu.
\end{eqnarray}
The conservation law ($\partial_\mu \,J^\mu_b = 0 $) can be proven by using the following EL-EoMs
\begin{eqnarray}
&&(\partial \cdot B) = 0, \quad    \Box \,\beta = 0, \quad \partial_\mu F^{\mu\nu} + \partial^\nu B = 0, \;\;
\varepsilon^{\mu\nu\sigma} \, \partial_\mu {\cal B} + \left(\partial^{\nu} B^{\sigma} - \partial^{\sigma} B^{\nu}\right) = 0, \nonumber\\
&& \Box \,C = 0, \quad \partial_\mu \left(\partial^{\mu} \bar C^{\nu} - \partial^{\nu} \bar C^{\mu}\right) - \partial^\nu \rho = 0, \quad
\partial_\mu \left(\partial^{\mu} C^{\nu} - \partial^{\nu} C^{\mu}\right) + \partial^\nu \lambda = 0,
\end{eqnarray}
that emerge out from the BRST-invariant Lagrangian density ${\cal L}_{B} $. Thus, our observations in (18), (19) and  use of the EL-EoMs in (20)
ensure that we have derived the expression for the conserved ($\partial_\mu \,J^\mu_b = 0 $) Noether BRST current($J^\mu_b $).
For our present combined system of the 3D free Abelian 1-form and 2-form free gauge theories, the expression for the conserved Noether 
BRST charge $Q_b$ is as follows:
\begin{eqnarray}
Q_b = \int d^2 x \,J^0_b &\equiv& 
\int d^2 x\, \Bigl [\left(\partial^{0} \bar C^{i} - \partial^{i} \bar C^{0}\right)\, \partial_i \beta - B \, \dot C 
- F^{0i} \, \partial_i C  - \rho \, \dot \beta \nonumber\\
&-&  \left(\partial^{0}  C^{i} - \partial^{i}  C^{0}\right) \, B_i - \varepsilon^{0ij}\, {\cal B} \, \partial_i C_j - \lambda \, B^0 \Bigr ].
\end{eqnarray}
It is crystal clear that the above charge is derived from the conserved Noether current (19).

We end this subsection with the following concluding remarks. First of all, we note that the Noether conserved BRST charge (21) is the generator for the
off-shell nilpotent BRST symmetry transformations (17) provided we compute the canonical conjugate momenta w.r.t. {\it all} the dynamical fields of 
our theory [cf. Eq. (16)] and exploit the theoretical strength of the canonical (anti)commutators in the generalized form of equation (11)
(where {\it only} the canonical commutator exists), namely;
\begin{eqnarray}
s_b \, \Phi (\vec{x}, t) =  \Big [ \Phi(\vec{x}, t), \; Q_b\Big]_{(\mp)}.
\end{eqnarray}
In the above, the generic field $\Phi (\vec{x}, t)$ stands for the bosonic as well as the fermionic fields of (16) and the subscript
$(\mp)$ on the square bracket (on the r.h.s) denotes the bracket to be (i) a commutator for  the generic field $\Phi (\vec{x}, t)$ being a bosonic
field, and  (ii) an anticommutator for the the generic field $\Phi (\vec{x}, t)$ being a fermionic field. Second, we point out that 
the kinetic terms (owing their origins to the exterior derivative of differential geometry) of the 
combined system of the 3D Abelian 1-form and 2-form theories remain
invariant [i.e. $s_b F_{\mu\nu} = 0, \, s_b H_{012} \equiv \frac{1}{2}\, \varepsilon^{\mu \nu \sigma} s_b (\partial_{\mu} B_{\nu \sigma}) = 0 $] 
under the BRST symmetry transformations ($s_b$). Finally, we observe that the Noether BRST charge $Q_b$ is {\it not} nilpotent 
(i.e. $ Q_b^2 \neq 0$) of order two even though it is computed from the off-shell nilpotent (i.e. $s_b^2 = 0$) version of the
BRST transformations (17). In other words, we note that the following is {\it true}, namely;
\begin{eqnarray}
s_b Q_b = -i \{ Q_b, \; Q_b \} \equiv \int d^2 x \bigl [ -\; \left(\partial^{0} B^{i} - \partial^{i} B^{0}\right)\, \partial_i \beta 
\bigr ] \neq 0, 
\end{eqnarray}
where the l.h.s. of the above relationship has been computed explicitly using the transformations (17) and the 
expression for the BRST charge  (21). In the above verification of the non-nilpotent nature of the {\it Noether} BRST charge, we have
also exploited the beauty and general nature of the relationship between the continuous symmetry transformation(s)
and their generator(s) as the Noether conserved charge(s). To be precise, we point out that we have proven, in equation (23), that:
$-i \,\{ Q_b, \; Q_b \} = -\, 2\,i\,  Q_b^2 \neq 0 $ which establishes the non-nilpotent 
(i.e. $ Q_b^2 \neq 0$)  nature of the conserved Noether BRST charge $Q_b$.

\subsection{Nilpotent Version of the BRST Charge}

The purpose of this subsection is to address the question of the derivation of the off-shell nilpotent 
BRST charge from the non-nilpotent  {\it Noether} BRST charge. The {\it nilpotency} property
of the BRST charge is important from the points of view of the mathematical as well as physical aspects of the BRST formalism. In other words, 
mathematically 
(i) to understand the BRST cohomology, and (ii) to establish the fermionic nature of the BRST charge, it is very essential to derive the
off-shell nilpotent ($Q_B^2 = 0$) version of the BRST charge $Q_B$ from the non-nilpotent (i.e. $Q_b^2 \neq 0$)
version of the Noether BRST charge $Q_b$
without spoiling the property of the conservation law. As far as the physical aspect is concerned, we demand that the physical states
(i.e. $|phys> $) of the {\it total} quantum Hilbert space of states (in the case of the BRST-quantized theory) are 
{\it those} that are  annihilated by
the conserved and {\it nilpotent} version of the BRST charge $Q_B$. The {\it latter} property is crucial to have consistency with the 
 Dirac quantization conditions for the systems that are endowed with constraints. In fact, we discuss {\it this} crucial issue
{\it clearly}  at the fag end of  our present subsection where we show that the physicality criterion (i.e. $Q_B \, |phys> = 0$)
w.r.t. the nilpotent version of the BRST charge leads to the annihilation of the physical states  (i.e. $|phys> $) by
the operator forms of the first-class constraints of the {\it classical} 3D combined theory of the free Abelian 1-form and 2-form  gauge fields.

 Against the backdrop of the above paragraph, it is pertinent to point out that, in our earlier work [34], we have provided the theoretical arguments for the 
 systematic derivation of the nilpotent version of the BRST charge from the non-nilpotent version of the {\it Noether} BRST charge
 where we have exploited {\it mainly} the beauty and strength of  (i) the EL-EoMs that are derived from the BRST-invariant Lagrangian density, and (ii)
 the Gauss divergence theorem due to which we drop the total {\it space} derivative terms. The above two inputs do {\it not} spoil the conservation law
 of the {\it modified} version of the Noether BRST charge. In addition to the above {\it two} inputs, we have also taken into account the 
theoretical strength of the BRST symmetry  transformations at {\it appropriate}
 places to ensure the perfect nilpotency property\footnote{In view of our observation in (23), we plan to prove that: $s_b Q_B = -i \,\{ Q_B, \; Q_B \} = 0 $
 which implies the off-shell nilpotency ($Q_B^2 = 0$) of the BRST charge $Q_B$ [cf. Eq. (33) below]. In our approach, the emphasis will be laid
on the explicit computation of $s_b Q_B = 0$ by using (17) and (32).} 
 of the BRST charge $Q_B$. Keeping in mind the above inputs, we focus on the following term of $Q_b$
 which  can be re-expressed, due to Gauss's divergence theorem, as 
\begin{eqnarray} 
\int d^2 x\, \bigl [-\, \varepsilon^{0ij}\, {\cal B} \, \partial_i C_j \big] &=&  \int d^2 x\,  \partial_i\,
\bigl [-\, \varepsilon^{0ij}\, {\cal B} \,  C_j \big] +
\int d^2 x\, \bigl [\, \varepsilon^{0ij}\, (\partial_i \,{\cal B})  \,  C_j \big] \nonumber\\
&\equiv& \int d^2 x\, \bigl [\, \varepsilon^{0ij}\, (\partial_i \,{\cal B})  \,  C_j \big],
\end{eqnarray}
because the {\it first} term on the r.h.s. does {\it not} contribute anything as both the fields, in the square-bracket, go
to zero as $x \to \pm \infty$. Using the following EL-EoM [cf. Eq. (20)]
\begin{eqnarray}
\varepsilon^{\mu\nu\sigma} \, \partial_\mu {\cal B} + \left(\partial^{\nu} B^{\sigma} - \partial^{\sigma} B^{\nu}\right) = 0
\qquad \Longrightarrow \qquad \varepsilon^{i0j} \, \partial_i {\cal B} + \left(\partial^{0} B^{j} - \partial^{j} B^{0}\right) = 0,
\end{eqnarray}
we can re-express the final form of the r.h.s. of (24) as follows
\begin{eqnarray} 
\int d^2 x\, \bigl [\, \varepsilon^{0ij}\, (\partial_i \,{\cal B})  \,  C_j \big] = \int d^2 x\, 
\bigl [(\partial^0 B^i - \partial^i B^0) \,  C_i \big],
\end{eqnarray}
which will be present as a part of the explicit expression for the nilpotent version (i.e. $Q_B^2 = 0$)
of the BRST charge $Q_B$.  At this juncture, 
as per the rules laid down in our earlier work [34], we apply the BRST symmetry transformation [cf. Eq. (17)]  on the r.h.s. of the above
equation which yields the following explicit expression:
\begin{eqnarray} 
s_b \, \Big [\int d^2 x\, 
\bigl [(\partial^0 B^i - \partial^i B^0) \,  C_i \big] \Big ] = -\, \int d^2 x\, 
\bigl [(\partial^0 B^i - \partial^i B^0) \,  \partial_i \beta \big].
\end{eqnarray}
We have to modify the appropriate term of the non-nilpotent version of the 
Noether BRST charge $Q_b$ [cf. Eq. (21)] so that when we apply the BRST symmetry transformation
on {\it a part} of the modified term, it cancels out with (27). In this connection, we perform the following modification in the 
first term of (21), namely;
\begin{eqnarray}  
\int d^2 x\, \left(\partial^{0} \bar C^{i} - \partial^{i} \bar C^{0}\right)\, \partial_i \beta =
 2\,\int d^2 x\, \left(\partial^{0} \bar C^{i} - \partial^{i} \bar C^{0}\right)\, \partial_i \beta - 
\int d^2 x\, \left(\partial^{0} \bar C^{i} - \partial^{i} \bar C^{0}\right)\, \partial_i \beta.
\end{eqnarray} 
It can be readily checked that if we apply the BRST transformation $s_b$ on the {\it second} term on the r.h.s. of the above
equation, it cancels out with (27). Hence, we have already obtained  {\it two} terms of the nilpotent version of the BRST charge $Q_B$ which 
are nothing but the sum of the r.h.s of (26) and the  {\it second} term on the r.h.s. of  (28).

The stage is now set  to concentrate on the first term of the r.h.s. of (28) which is present inside the integral [cf. Eq. (21)] as follows:
\begin{eqnarray}  
 2\, \int d^2 x\, \left(\partial^{0} \bar C^{i} - \partial^{i} \bar C^{0}\right)\, \partial_i \beta &=& 
 2\, \int d^2 x\, \partial_i \big [\left(\partial^{0} \bar C^{i} - \partial^{i} \bar C^{0}\right)\,  \beta \big ] \nonumber\\
 &-& 2\, \int d^2 x\, \partial_i \big [\left(\partial^{0} \bar C^{i} - \partial^{i} \bar C^{0}\right) \big ]\,  \beta. 
\end{eqnarray} 
It is clear that the {\it first} term on the r.h.s. of the above equation will drop out due to Gauss's divergence theorem. As far as the
second term of (29) on the r.h.s. is concerned, we exploit the beauty and strength of the following EL-EoM (with the choice $\nu = 0 $), namely;
\begin{eqnarray}
\partial_\mu \left(\partial^{\mu} \bar C^{\nu} - \partial^{\nu} \bar C^{\mu}\right) - \partial^\nu \rho = 0\quad \Longrightarrow \quad
\partial_i \left(\partial^{i} \bar C^{0} - \partial^{0} \bar C^{i}\right) = \dot \rho,
\end{eqnarray}
to re-express the {\it second} term on the r.h.s. of (29) as follows:
\begin{eqnarray}
 - 2\, \int d^2 x\, \partial_i \big [\left(\partial^{0} \bar C^{i} - \partial^{i} \bar C^{0}\right) \big ]\,  \beta = 
 +\, 2\, \int d^2 x\, \partial_i \big [\left(\partial^{i} \bar C^{0} - \partial^{0} \bar C^{i}\right) \big ]\,  \beta  \equiv 
 2\, \int d^2 x\, \dot \rho\, \beta. 
\end{eqnarray}
It is straightforward to check that if we apply the BRST symmetry transformation $s_b$ on the above {\it final} 
expression, it turns out to be zero (i.e. $s_b [\dot \rho\, \beta ] = 0 $). Thus, the precise expression for the off-shell nilpotent 
version ($Q_B^2 = 0$)
of the BRST charge $Q_B$ (that emerges out from the non-nilpotent $Q_b^2 \neq 0$ version of the 
Noether charge $Q_b$) is:
\begin{eqnarray}
Q_b \to Q_B &=& 
\int d^2 x\, \Bigl [ \left(\partial^{0} B^{i} - \partial^{i} B^{0}\right)\, C_i -
\left(\partial^{0} \bar C^{i} - \partial^{i} \bar C^{0}\right)\, \partial_i \beta  + 2\, \dot \rho\, \beta  - \rho \, \dot \beta
 \nonumber\\
&-&  B \, \dot C - \lambda \, B^0 -  \left(\partial^{0}  C^{i} - \partial^{i}  C^{0}\right) \, B_i  - F^{0i} \, \partial_i C  \Bigr ].
\end{eqnarray}
It is worthwhile to mention that {\it only} the first three terms, in the above expression for the BRST charge $Q_B$, are {\it new}  that
have been obtained by exploiting the theoretical tricks that have been proposed in our earlier work [34]. To be precise,
we have used mainly the Gauss divergence theorem and the appropriate set of EL-EoMs to derive the above three
{\it new} terms. The rest of the terms of $Q_B$ are {\it same} as the ones that are present in the expression for 
the Noether BRST charge $Q_b$ because
{\it these} terms are BRST-invariant. It is interesting to point out that, ultimately, we can check that the following is 
{\it true}, namely;
\begin{eqnarray}
s_b Q_B = -i\, \{ Q_B, \; Q_B \} = 0, \quad \Longrightarrow \quad Q_B^2 = 0.
\end{eqnarray}
To corroborate the above claim, we have to explicitly compute the l.h.s. of the above equation by directly applying the off-shell nilpotent 
BRST symmetry transformations $s_b$ [cf. Eq. (17)] on the above expression for $Q_B$ [cf. Eq. (32)]. It is pertinent to point out that
{\it both} the charges (21) and (32) are (i) equivalent to each-other, and (ii) conserved. This is due to the fact that we have used
{\it only} (i) the appropriate EL-EoMs, and (ii) the Gauss divergence theorem to derive (32) from (21). In the context of Noether's theorem,
it is an undeniable truth that the Noether conserved charge (corresponding to a continuous symmetry transformation) can be recast
into many different forms by using the appropriate EL-EoMs and the Gauss divergence theorem (without spoiling the conservation law).

We end this subsection by establishing the connection between the physicality criterion (i.e. $Q_B \, |phys> = 0 $)
w.r.t. the nilpotent BRST charge $Q_B$  and the first-class constraints of the {\it classical} gauge theory (cf. Sec. 2 for details). In this
connection, it is pertinent to point out that, right from the beginning, we note that the FP-ghost fields are {\it decoupled} from
the rest of the theory. As a consequence, quantum Hilbert space of states (of the BRST-quantized theory) 
is the direct product (see, e.g. [35]) 
of the physical states and the ghost states. The {\it latter} states are operated {\it only} by the ghost fields 
(with the non-zero ghost numbers) and the former states are operated {\it only} by the physical fields (with zero ghost number). A close look
at the nilpotent version of the BRST charge (32) shows that each term of its expression contains, at least, one ghost field (and the ghost number 
of each term is + 1).  Hence, when the BRST charge $Q_B$ operates on a quantum state, the ghost fields (with non-zero ghost numbers) will
operate on the ghost states and produce the non-zero results. To satisfy, the requirement (i.e. $Q_B \, |phys> = 0 $), we have to look for 
the fields with zero ghost number that are present in the expression for $Q_B$. These {\it zero} ghost number fields\footnote{It is worthwhile
to point out that we have {\it not} taken into account $B^0 \, |phys> = 0$ as a condition on the physical state ($|phys> = 0 $)
 because the field $B^0$ is associated with $\lambda$ which is {\it not} the basic ghost field of our theory. Moreover, a close
 look at the Lagrangian density ${\cal L}_B$ [cf. Eq. (16)] demonstrates that the field $B^0$ is the conjugate momentum w.r.t.  the 
scalar field $\phi$ and 
 the {\it former} is {\it not} a constraint on our theory.}
 will operate on the physical states (i.e. $|phys> $) to yield the zero result (in $Q_B \, |phys> = 0 $). Thus, we obtain the following
(from  $Q_B \, |phys> = 0 $), namely; 
\begin{eqnarray}
&& B_i \, |phys> = 0, \qquad \big (\partial^0 B^i - \partial^i B^0 \big )\, |phys> = 0, \nonumber\\
&& B \, |phys> = 0, \qquad \big (\partial_i F^{0i}\big )\, |phys> \equiv \dot B\, |phys> = 0.
\end{eqnarray}
The above conditions correspond to the physicality criterion w.r.t. the conserved and  nilpotent version of the BRST charge $Q_B$. In the above equation (34),
we have used the Gauss divergence theorem to re-express: $\int d^2 x\, (-\, F^{0i} \, \partial_i C) = + \int d^2 x\, (\partial_i F^{0i}) \,C $
and used the EL-EoM: $\partial_\mu F^{\mu\nu} + \partial^\nu B = 0 $ (with the choice $\nu = 0$) to obtain the last entry of the
above equation. To establish the connection between the quantum conditions in (34) with the first-class
constraints of our {\it classical} gauge theory (cf. Sec. 2 for details), we have to focus on the BRST-invariant Lagrangian density 
${\cal L}_B$ [cf. Eq. (16)] and note the following
\begin{eqnarray} 
&&\Pi^0_{(A)}   = \frac{\partial\, {\cal L}_{B}}{\partial\, (\partial_0\, A_0)} = -\, F^{00} - \,\eta^{00} B \equiv -\, B, \qquad
\Pi^i_{(A)}   = -\, F^{0i} \equiv E_i
   \nonumber\\ 
&&\Pi_{(B)}^{0i}  = \frac {\partial {\cal L}_{B}}{\partial (\partial_0 B_{0i})}
 = \frac{1}{2}\, 
 \varepsilon^{00i}\, {\cal B} +  \frac{1}{2}\, B^i \equiv - \, \frac{1}{2}\, B_i, \;\;\qquad
\Pi_{(B)}^{ij} = \frac{1}{2}\, \varepsilon^{0ij}\, {\cal B},  
\end{eqnarray}
which demonstrate that the primary constraints (i.e. $\Pi^0_{(A)} \approx 0, \; \Pi_{(B)}^{0i} \approx 0 $) of the classical
gauge theory are traded with the Nakanishi-Lautrup auxiliary fields (e.g. $B,\, B_i$)
of the BRST-quantized theory. Thus, the first and third entries of equation
(34) correspond to the annihilation of the physical states (i.e. $|phys> $), at the quantum level, by the operator forms of the
primary constraints of the {\it classical} gauge theory. On the other hand, the last entry in (34) implies the
annihilation of the physical states by the operator form of the secondary constraint ($\partial_i \Pi^i_{(A)} \approx 0 $),
corresponding to the free Abelian 1-form gauge theory, in the language of the time derivative (i.e. $\dot B $) on the 
Nakansihi-Laurtup auxiliary field. Let us focus on the second entry 
of (34) now. It is clear from the EL-EoM (25) that: $\big (\partial^0 B^i - \partial^i B^0 \big ) = \varepsilon^{0ji}\, \partial_j {\cal B}
\equiv 2 \, \partial_j \Pi^{ji}_{(B)}$ [cf. Eq. (35)]. Thus, it is pretty obvious that we have obtained the annihilation of the physical states (i.e. $|phys> $)
by the secondary constraint (i.e. $\partial_j \Pi^{ji}_{(B)} \approx 0$) of the {\it classical } version of the Abelain 2-form
gauge theory. This SC  is expressed precisely in terms of the 
specific combination of derivatives [i.e. $\big (\partial^0 B^i - \partial^i B^0 \big ) $] on the
Nakanishi-Lautrup auxiliary fields at the level of the BRST-quantized theory.\\

\section{Anti-BRST invariant Lagrangian Density: Off-Shell Nilpotent Anti-BRST Symmetries}

This section is divided into two subsections. In Subsec. 4.1, we derive the Noether conserved anti-BRST charge 
from the off-shell nilpotent anti-BRST symmetries and show that
it is {\it not} off-shell nilpotent. Our Subsec. 4.2 is devoted to the derivation of the off-shell nilpotent version of the 
anti-BRST charge from the non-nilpotent {\it Noether} anti-BRST charge.

\subsection{Anti-BRST Symmetries and Noether Anti-BRST Charge}

Analogous to the Lagrangian density ${\cal L}_B$ [cf. Eq. (16)], 
the classical Lagrangian density (1) can be generalized to the quantum level where we have an anti-BRST invariant Lagrangian density
$({\cal L}_{\bar B})$  that incorporates into itself the gauge-fixing and FP-ghost terms  as 
\begin{eqnarray}
{\cal L}_{\bar B} &=& \mathcal{B} \left(\frac{1}{2} \varepsilon^{\mu \nu \sigma} \partial_{\mu} B_{\nu \sigma}\right)-\frac{\mathcal{B}^{2}}{2}
+ \bar B^{\mu} \left(\partial^{\nu} B_{\nu \mu} + \partial_{\mu} \phi \right)
- \frac{\bar B^{\mu} \bar B_{\mu}}{2} \nonumber\\
&-& \frac{1}{4} F^{\mu \nu} F_{\mu \nu} + \frac{B^{2}}{2}   - B \, (\partial \cdot A) - \partial_{\mu} \bar{C} \, \partial^{\mu} C 
+ \partial_{\mu} \,  \bar{\beta}\, \partial^{\mu} \beta \nonumber\\
&+&\left(\partial_{\mu} \bar{C}_{\nu}
 -\partial_{\nu} \bar{C}_{\mu}\right)\left(\partial^{\mu} C^{\nu}\right)  
  +(\partial \cdot \bar{C}+\rho)\,  \lambda+(\partial \cdot C-\lambda)\, \rho, 
\end{eqnarray}
where $\bar B_\mu$ is a new Nakanishi-Lautrup type auxiliary field that has been invoked to 
linearize  the gauge-fixing term for the Abelian antisymmetric tensor gauge field $B_{\mu\nu}$ where we have taken into account 
$\phi \rightarrow - \,\phi$ for the sake of generality of the gauge-fixing term. 
The above Lagrangian density $({\cal L}_{\bar B})$ respects the following infinitesimal, continuous and off-shell nilpotent 
($s_{ab}^2 = 0$) version of the anti-BRST symmetry transformations ($s_{ab}$)
\begin{eqnarray}
&& s_{ab} B_{\mu \nu}=-\left(\partial_{\mu} \bar C_{\nu} - \partial_{\nu} \bar C_{\mu}\right), \quad s_{ab} \bar C_{\mu}= -\,\partial_{\mu} \bar \beta, 
\quad s_{ab} {C}_{\mu}= \bar B_{\mu}, \quad s_{ab} C = -\, B,
\nonumber\\
&& s_{ab} A_{\mu}= \partial_{\mu} \bar C, \;\; s_{ab} {\beta} = - \lambda, \;\; s_{ab} \phi = \rho, \;\;
 s_{ab}\left[ \rho, \lambda, \bar C, \bar \beta,  B, \mathcal{B}, \bar B_{\mu},  F_{\mu\nu}, H_{\mu \nu \lambda}\right]=0,
\end{eqnarray}
because we observe that ${\cal L}_{\bar B}$ transforms to a total spacetime derivative, namely; 
\begin{eqnarray}
s_{ab} \mathcal{L}_{\bar B} = - \,\partial_{\mu} \Big[\left(\partial^{\mu} \bar C^{\nu}-\partial^{\nu} \bar C^{\mu}\right) \bar B_{\nu}
-\, \rho \,\bar B^{\mu}
+ \lambda\, \partial^{\mu} \bar \beta + B\,\partial^{\mu} \bar C\Big]. 
\end{eqnarray} 
As a consequence, the action integral $S = \int d^3 x \, {\cal L}_{\bar B}$  remains invariant (i.e. $s_{ab}\, S = 0$) due to the 
Gauss divergence theorem (because all the physical fields of our theory vanish off as $x \rightarrow \pm \infty$). 
According to Noether's theorem, whenever action integral remains invariant under a continuous symmetry transformation, there
is always a conserved current which is christened as the Noether current.  
As a consequence of our observation in (38), we have
the following explicit expression for the Noether current ($J^\mu_{ab}$)
\begin{eqnarray}
J^\mu_{ab} &=& \rho \, \bar B^\mu -
\left(\partial^{\mu} \bar C^{\nu} - \partial^{\nu} \bar C^{\mu}\right)\, \bar B_\nu - B\, \partial^\mu \bar C - F^{\mu\nu}\, \partial_\nu \bar C
 - \lambda \,\partial^\mu \bar \beta \nonumber\\
&-& \varepsilon^{\mu\nu\sigma} \, {\cal B} \, \partial_\nu \bar C_\sigma - \left(\partial^{\mu} C^{\nu} 
- \partial^{\nu} C^{\mu}\right)\,
\partial_\nu \bar \beta,
\end{eqnarray}
where the standard theoretical techniques of the Noether theorem have been exploited (in the above derivation).  
At this juncture, we exploit the theoretical beauty and strength of the following EL-EoMs that emerge out from the 
Lagrangian density ${\cal L}_{\bar B}$, namely;
\begin{eqnarray}
&&(\partial \cdot \bar B) = 0, \quad    \Box \,\bar \beta = 0, \quad \partial_\mu F^{\mu\nu} + \partial^\nu B = 0, \;\;
\varepsilon^{\mu\nu\sigma} \, \partial_\mu {\cal B} + \left(\partial^{\nu} \bar B^{\sigma} - \partial^{\sigma} \bar B^{\nu}\right) = 0, \nonumber\\
&& \Box \,\bar C = 0, \quad \partial_\mu \left(\partial^{\mu} \bar C^{\nu} - \partial^{\nu} \bar C^{\mu}\right) - \partial^\nu \rho = 0, \quad
\partial_\mu \left(\partial^{\mu} C^{\nu} - \partial^{\nu} C^{\mu}\right) + \partial^\nu \lambda = 0,
\end{eqnarray}
 to prove the conservation law: $\partial_\mu \, J^\mu_{ab} = 0$. This 
anti-BRST Noether current (in particular its zeroth component)
 leads to the definition of the  anti-BRST charge $Q_{ab} = \int d^2 x \, J^0_{ab}$  as
\begin{eqnarray}
Q_{ab} &=&  
\int d^2 x\, \Bigl [\rho \, \bar B^0 - \left(\partial^{0} \bar C^{i} - \partial^{i} \bar C^{0}\right)\, \bar B_i - B \, \dot {\bar C} 
- F^{0i} \, \partial_i \,\bar C  - \lambda \, \dot {\bar \beta} \nonumber\\
&-&  \left(\partial^{0}  C^{i} - \partial^{i}  C^{0}\right) \, \partial_i \,\bar \beta - \varepsilon^{0ij}\, {\cal B} \, \partial_i\, \bar C_j  \Bigr ],
\end{eqnarray}
which is found to be the generator for the off-shell nilpotent anti-BRST symmetry transformations (37).  To prove it, we have to use
the standard relationship between the infinitesimal, continuous and off-shell nilpotent anti-BRST symmetry transformations (37) and the
above conserved anti-BRST Noether charge. In other words, we have to use 
the general and standard relationship (22) with the replacements: $s_b \to s_{ab}, \; Q_b \to Q_{ab} $.

We conclude this subsection with the following clinching remarks.  First, we note that, just like BRST symmetry transformations, the anti-BRST
symmetries leave the kinetic terms (owing their origins to the exterior derivative of differential geometry) 
of the 3D Abelian 1-form and 2-form theories invariant 
[i.e. $s_{ab} F_{\mu\nu} = 0, \, s_{ab} H_{012} \equiv \frac{1}{2}\, \varepsilon^{\mu \nu \sigma} s_{ab} (\partial_{\mu} B_{\nu \sigma}) = 0 $].
Second, we would like to lay emphasis on the fact that the (anti-)BRST invariant Lagrangian densities $ {\cal L}_{\bar B}$ and
$ {\cal L}_{B}$ are the equivalent {\it quantum} generalizations of the classical Lagrangian density $ {\cal L}^{(3D)}_{(0)}$ [cf. Eq. (14)] 
on the submanifold of the fields where the CF-type restriction (i.e. $B_\mu - \bar B_\mu + 2\, \partial_\mu \phi = 0$)
is satisfied (cf. Appendix A below). Third, we point out that the EL-EoMs w.r.t. the Nakanishi-Lautrup auxiliary fields $B_\mu$ and  $\bar B_\mu$
from the Lagrangian densities $ {\cal L}_{B}$ and $ {\cal L}_{\bar B}$, respectively, lead to the following relationship
\begin{eqnarray}
B_\mu = \partial^\nu B_{\nu\mu} - \partial_\mu \phi, \qquad \bar B_\mu = \partial^\nu B_{\nu\mu} + \partial_\mu \phi \qquad
\Longrightarrow  \qquad B_\mu - \bar B_\mu + 2\, \partial_\mu \phi = 0,
 \end{eqnarray}
 which turns out to be the CF-type restriction on our theory (cf. Appendices A and B below). Fourth, for this CF-type restriction to be 
{\it physically} meaningful
 on our BRST-quantized theory, it should be (anti-)BRST invariant (i.e. $s_{(a)b} \bigl [B_\mu - \bar B_\mu + 2\, \partial_\mu \phi \bigr] = 0$).
 This requirement leads to the following {\it additional} (anti-)BRST transformations, namely; 
\begin{eqnarray}
s_b \bar B_\mu = 2\, \partial_\mu \lambda, \qquad s_{ab} B_\mu = -\,2\, \partial_\mu \rho,
\end{eqnarray}
which are also off-shell nilpotent [but these transformations are {\it not} present in (37) and (17)]. Fifth, the absolute anticommutativity
property (i.e. $\{ s_b, \; s_{ab} \} = 0 $)
 of the nilpotent
(anti-)BRST transformations is {\it true} for all the fields except the gauge field $B_{\mu\nu}$ and the Lorentz vector
(anti-)ghost fields $(\bar C_\mu)C_\mu$
 because we observe the following 
\begin{eqnarray}
&&\{ s_b, \; s_{ab} \} \, B_{\mu\nu} = \partial_\mu \big ( B_\nu - \bar B_\nu \big ) -  \partial_\nu \big ( B_\mu - \bar B_\mu \big ),\nonumber\\
&& \{ s_b, \; s_{ab} \} \, C_\mu = 3\, \partial_\mu \lambda, \qquad \{ s_b, \; s_{ab} \} \, \bar C_\mu = 3\, \partial_\mu \rho,
\end{eqnarray}
which turns out to be {\it zero} on the submanifold of the fields where the CF-type restriction (i.e. $B_\mu - \bar B_\mu + 2\, \partial_\mu \phi = 0$)
is satisfied in the case of the gauge field $B_{\mu\nu} $. However, as far as the vector (anti-)ghost fields $(\bar C_\mu)C_\mu$ are concerned,
we note that the anticommutativity property is satisfied {\it only} up to the  Abelian  U(1) symmetry-type transformations.
In other words, we have the validity of the absolute anticommutativity property of the (anti-)BRST transformations 
(i) due to the imposition  of the (anti-)BRST invariant CF-type restriction, and (ii)
modulo the Abelian  
U(1)  gauge symmetry-type transformations\footnote{The observations in (44) are {\it not} new as far as the (anti-)ghost fields are concerned. 
In [29], such kind of anticommutativity property has been discussed in the context of the BRST approach to the free Abelian 2-form gauge theory
where the nilpotent BRST and anti-BRST transformations have been shown to be anticommuting {\it only} up to the 
Abelian U(1) gauge symmetry-type transformations
in the ghost-sector.}. 
Sixth, it can be explicitly checked that the Noether anti-BRST charge $Q_{ab}$ is {\it not} off-shell nilpotent 
(i.e.  $Q^2_{ab} \neq 0$) of order two. In other words, we note the following explicit relationship
\begin{eqnarray}
s_{ab} Q_{ab} = -i \{ Q_{ab}, \; Q_{ab} \} \equiv 
\int d^2 x \bigl [ -\; \left(\partial^{0} \bar B^{i} - \partial^{i} \bar B^{0}\right)\, \partial_i \bar \beta 
\bigr ] \neq 0, 
\end{eqnarray}
which implies, ultimately, that the Noether conserved anti-BRST charge is non-nilpotent.  
Finally, the Lagrangian densities $ {\cal L}_{B}$ and $ {\cal L}_{\bar B}$
are {\it equivalent} w.r.t. the nilpotent (anti-)BRST symmetries provided we take into account the sanctity of the CF-type restriction. To be precise,
we note the following explicit transformations, namely;
\begin{eqnarray}
s_b {\cal L}_{\bar B} &=& \partial_\mu \Big [2\, (\partial_\nu B^{\nu\mu}) \, \lambda 
- \left(\partial^{\mu} C^{\nu}  - \partial^{\nu} C^{\mu}\right)\, \bar B_\nu - B \,\partial^\mu C  
- \rho \,\partial^\mu \beta - \lambda\, B^\mu \Big ] \nonumber\\
 &+& (\partial^\mu \lambda) \, \big [B_\mu - \bar B_\mu + 2\, \partial_\mu \phi \big ]
 - \left(\partial^{\mu} {C}^{\nu} - \partial^{\nu} {C}^{\mu}\right)\, \partial_\mu \big [B_\nu - \bar B_\nu + 2\, \partial_\nu \phi \big], 
\end{eqnarray}
\begin{eqnarray}
s_{ab} {\cal L}_{ B} &=& \partial_\mu \Big [\rho\, \bar B^\mu\,- 2\, (\partial_\nu B^{\nu\mu}) \, \rho - \left(\partial^{\mu} \bar {C}^{\nu}
 - \partial^{\nu} \bar {C}^{\mu}\right)\,  B_\nu - B \,\partial^\mu \bar C  - \lambda\, \partial^\mu \bar \beta  \Big ] \nonumber\\
 &+& (\partial^\mu \rho) \, \big [B_\mu - \bar B_\mu + 2\, \partial_\mu \phi \big] + \left(\partial^{\mu} \bar {C}^{\nu}
 - \partial^{\nu} \bar{C}^{\mu}\right)\, \partial_\mu \big [B_\nu - \bar B_\nu + 2\, \partial_\nu \phi \big], 
\end{eqnarray}
which establish that, in addition to our observations in (18) and (38), the Lagrangian densities $ {\cal L}_{B}$ and $ {\cal L}_{\bar B}$ respect
the anti-BRST and BRST symmetry transformations [cf.  Eqs. (47),(46)], respectively,  provided we exploit the validity of the 
(anti-BRST invariant CF-type restriction  (i.e. $B_\mu - \bar B_\mu + 2\, \partial_\mu \phi = 0$)  that is 
present on our theory. Thus, as far as the symmetry considerations are concerned, {\it both} the coupled Lagrangian densities  
$ {\cal L}_{B}$ and $ {\cal L}_{\bar B}$ are {\it equivalent} because both of them respect {\it both} the BRST as well as the anti-BRST symmetry
transformations on the submanifold of the fields where the 
(anti-)BRST invariant CF-type restriction (i.e. $B_\mu - \bar B_\mu + 2\, \partial_\mu \phi = 0$) is satisfied.

\subsection{Anti-BRST Charge: Off-Shell Nilpotent Version}

In view of our detailed discussion in our subsection 3.2 on the theoretical tricks to obtain the off-shell nilpotent version of the BRST charge 
($Q_B$) from the non-nilpotent Noether BRST charge ($Q_b$), we shall be very brief in the derivation of the off-shell nilpotent
version of the anti-BRST charge ($Q_{AB}$) from the non-nilpotent Noether anti-BRST charge ($Q_{ab}$). First of all, we focus on the
following term of the non-nilpotent version of the Noether anti-BRST charge [cf. Eq. (41)], namely;
\begin{eqnarray} 
\int d^2 x\, \bigl [-\, \varepsilon^{0ij}\, {\cal B} \, \partial_i \bar C_j \big] = \int d^2 x\, 
\bigl [\, \varepsilon^{0ij}\, (\partial_i \,{\cal B})  \, \bar  C_j \big],
\end{eqnarray}
where we have used the Gauss divergence theorem. Using the following EL-EoM that is derived from 
the anti-BRST invariant ${\cal L}_{\bar B}$ [cf. Eq. (36)], namely;
\begin{eqnarray}
\varepsilon^{\mu\nu\sigma} \, \partial_\mu {\cal B} + \left(\partial^{\nu} \bar B^{\sigma} - \partial^{\sigma} \bar B^{\nu}\right) = 0
\qquad \Longrightarrow \qquad \varepsilon^{i0j} \, \partial_i {\cal B} + \left(\partial^{0} \bar B^{j} - \partial^{j} \bar B^{0}\right) = 0,
\end{eqnarray}
we can recast the r.h.s. of the above equation as follows:
\begin{eqnarray} 
\int d^2 x\, \bigl [\, \varepsilon^{0ij}\, (\partial_i \,{\cal B})  \,  \bar C_j \big] = \int d^2 x\, 
\bigl [(\partial^0 \bar B^i - \partial^i \bar B^0) \,  \bar C_i \big].
\end{eqnarray}
The above expression will be  a part of the off-shell nilpotent (i.e. $Q^2_{AB} = 0$)
version of the anti-BRST charge $Q_{AB}$. As per the proposal suggested in our
earlier work [34], we have to apply the anti-BRST symmetry transformations ($s_{ab}$) on the above  expression to obtain:
\begin{eqnarray} 
 -\, \int d^2 x\, 
\bigl [(\partial^0 \bar B^i - \partial^i \bar B^0) \,  \partial_i \bar \beta \big].
\end{eqnarray}
At this stage, we have to modify an appropriate term of the non-nilpotent version of the anti-BRST charge $Q_{ab}$ [cf. Eq. (41)] such
that when we apply the anti-BRST symmetry transformations on a part of this modified term, it should cancel out with (51). This
modification for the {\it sixth} term of $Q_{ab}$ [cf. Eq. (41)] 
is as follows:
\begin{eqnarray}  
-\,\int d^2 x\, \left(\partial^{0}  C^{i} - \partial^{i}  C^{0}\right)\, \partial_i \bar \beta &=& 
-\, 2\, \int d^2 x\, \left(\partial^{0} C^{i} - \partial^{i}  C^{0}\right)\, \partial_i \bar \beta \nonumber\\
&+&
\int d^2 x\, \left(\partial^{0}  C^{i} - \partial^{i}  C^{0}\right)\, \partial_i \bar \beta.
\end{eqnarray}
It is crystal clear that if we apply  $s_{ab}$ [cf. Eq. (37)] on the {\it second} term on the r.h.s. of the above equation, we find
that it cancels out with (51). Hence, we have obtained {\it two} terms of the nilpotent (i.e. $Q_{AB}^2 = 0 $) version of the anti-BRST charge
$Q_{AB}$ as follows:
\begin{eqnarray} 
\int d^2 x\, 
\Bigl [(\partial^0 \bar B^i - \partial^i \bar B^0) \,  \bar C_i  
+ \left(\partial^{0}  C^{i} - \partial^{i}  C^{0}\right)\, \partial_i \bar \beta.
 \Big].
\end{eqnarray}
A  close look at the above equation demonstrate that it is nothing but the sum of (50) and the second term on the r.h.s. of (52). Now
we concentrate on the {\it first} term of (52) which can be expressed, using the Gauss divergence theorem, as 
\begin{eqnarray}  
-\, 2\, \int d^2 x\, \left(\partial^{0} C^{i} - \partial^{i}  C^{0}\right)\, \partial_i \bar \beta  = 
&-&\, 2\,
\int d^2 x\, \partial_i \left(\partial^{i}  C^{0}  - \partial^{0}  C^{i}  \right)\, \bar \beta.
\end{eqnarray}
At this juncture, we exploit the theoretical strength of the following EL-EoM 
\begin{eqnarray}
\partial_\mu \left(\partial^{\mu} C^{\nu} - \partial^{\nu}  C^{\mu}\right) + \partial^\nu \lambda = 0\quad \Longrightarrow \quad
\partial_i \left(\partial^{i}  C^{0} - \partial^{0}  C^{i}\right) =  -\,\dot \lambda,
\end{eqnarray}
to re-express the r.h.s. of (54) as follows:
\begin{eqnarray}
 - 2\, \int d^2 x\, \partial_i \big [\left(\partial^{0}  C^{i} - \partial^{i}  C^{0}\right) \big ]\,  \bar \beta = 
 - \, 2\, \int d^2 x\, \partial_i \big [\left(\partial^{i}  C^{0} - \partial^{0}  C^{i}\right) \big ]\,  \bar \beta  \equiv 
 2\, \int d^2 x\, \dot \lambda\, \bar \beta. 
\end{eqnarray}
It is worthwhile to point out that, in the EL-EoM (55), we have made the choice $ \nu = 0$ and  we observe that the final form of (56) is an
anti-BRST invariant (i.e. $s_{ab} [2\, \int d^2 x\, \dot \lambda\, \bar \beta] = 0 $) quantity. Thus, the final form of the
off-shell nilpotent version of the anti-BRST charge $Q_{AB}$, derived from the non-nilpotent version $Q_{ab}$, is as follows
\begin{eqnarray}
Q_{ab} \to Q_{AB} &=& 
\int d^2 x\, \Bigl [ \left(\partial^{0} \bar B^{i} - \partial^{i} \bar B^{0}\right)\, \bar C_i +
\left(\partial^{0}  C^{i} - \partial^{i}  C^{0}\right)\, \partial_i \bar \beta  + 2\, \dot \lambda\, \bar \beta  - \lambda \, \dot {\bar \beta}
 \nonumber\\
&-&  B \, \dot {\bar  C} + \dot B \, \bar C + \rho \, \bar B^0 -  \left(\partial^{0} \bar  C^{i} - \partial^{i}  \bar C^{0}\right) \, \bar B_i  \Big ],
\end{eqnarray}
where we have already used: $- \int d^2 x\, [F^{0i}\, \partial_i \bar C] =  + \int d^2 x \,[(\partial_i F^{0i}) \, \bar C] \equiv 
 \int d^2 x \,\dot B \, \bar C$ due to the application of the EL-EoM: $\partial_\mu \, F^{\mu\nu} + \partial^\nu B = 0 $ which
 implies that: $ (\partial_i F^{0i}) = \dot B$. It is straightforward  now to check that 
\begin{eqnarray}
s_{ab} Q_{AB} = -i\, \{ Q_{AB}, \; Q_{AB} \} = 0, \quad \Longrightarrow \quad Q_{AB}^2 = 0,
\end{eqnarray}
where the l.h.s. can be precisely computed by the direct application of 
the anti-BRST symmetry  transformations (37) on the explicit expression for $Q_{AB}$ in (57).

We conclude this subsection with the following remarks. First of all, we note that the anti-BRST charge 
(i.e. $\int d^2 x\, [\dot B \, \bar C  - B \, \dot {\bar  C} ]$) for the Abelian 1-form gauge theory 
does {\it not} create any problem in the proof of the nilpotency property as it is very simple theory
(with a {\it trivial} CF-type restriction).  Second, we find that the Noether anti-BRST
charge (41) is found to be non-nilpotent because of the existence of the {\it non-trivial} CF-type
restriction $B_\mu - \bar B_\mu + 2\, \partial_\mu \phi = 0$ in the BRST quantization of the
Abelian 2-form theory. Third, we observe that the physicality criterion (i.e. $Q_{AB} \, |phys> = 0 $)
w.r.t. the nilpotent version of the anti-BRST charge $Q_{AB}$ produces {\it exactly} the same conditions on the
physical states (i.e. $|phys> $) as we have obtained in (34) w.r.t. the nilpotent version of the
BRST charge. To be precise, we obtain the following from $Q_{AB} \, |phys> = 0 $, namely;
\begin{eqnarray}
&& \bar B_i \, |phys> = 0, \qquad \big (\partial^0 \bar B^i - \partial^i \bar B^0 \big )\, |phys> = 0, \nonumber\\
&& B \, |phys> = 0, \qquad \big (\partial_i F^{0i}\big )\, |phys> \equiv \dot B\, |phys> = 0,
\end{eqnarray}
where there has been {\it only}  change of the auxiliary fields:  $B_i \to \bar B_i, \; B_0 \to \bar B_0 $. This does {\it not} 
lead to any new physics because, we point out that, the operator forms of the first-class constraints 
{\it still} annihilate the physical states (i.e. $|phys> $). The arguments will go along similar lines
as we have {\it already} done, after equation (34), in the context of the physicality criterion (i.e. $Q_{B} \, |phys> = 0 $)
w.r.t. the conserved and nilpotent  (i.e. $Q^2_B = 0$) version of the BRST charge $Q_B$ in the subsection 3.2. 


\section{(Anti-)co-BRST Symmetries: Conserved Charges}

In addition to the off-shell nilpotent (i.e. $s_{(a)b}^2 = 0$) and absolutely anticommuting
(i.e. $s_b s_{ab} + s_{ab} s_b = 0 $) (anti-)BRST symmetry transformations [$s_{(a)b} $], there is another set of
off-shell nilpotent (i.e. $s_{(a)d}^2 = 0$) (anti-)dual-BRST [or (anti-)co-BRST] symmetry transformations [$s_{(a)d} $]
that are respected by {\it both} the Lagrangian densities ${\cal L}_B$ and  ${\cal L}_{\bar B}$ [cf. Eqs. (16),(36)].
To be precise, we observe that under the following infinitesimal, continuous and off-shell nilpotent
(anti-)dual BRST symmetry transformations [$s_{(a)d} $], namely;
\begin{eqnarray}
&&s_{ad}\, B_{\mu\nu} = -\, \varepsilon_{\mu\nu\sigma} \, \partial^\sigma\, C, \qquad
s_{ad}\, C_\mu =\, \partial_\mu\, \beta, \qquad
s_{ad}\, \bar C = {\cal B},  \quad s_{ad} \, \bar \beta =\, \rho, \nonumber\\
&& s_{ad}\, \Big[\rho ,\, \lambda, \, C, \, \beta, \,  B, \, {\cal B}, \,
\phi, \,\bar C_\mu, 
(\partial^\nu\, B_{\nu\mu}), \,  A_\mu,    B_{\mu}, \, \bar B_{\mu}, \, F_{\mu\nu}\Big] = 0, 
\end{eqnarray}
\begin{eqnarray}
&&s_{d}\, B_{\mu\nu} = -\, \varepsilon_{\mu\nu\sigma} \, \partial^\sigma\, \bar C, \qquad
s_{d}\, \bar C_\mu = -\, \partial_\mu\, \bar \beta, \qquad
s_{d}\, C = -\, {\cal B},  \quad
s_{d}\, \beta = -\, \lambda,  \nonumber\\
&&s_{d}\, \Big[\rho ,\, \lambda, \, \bar C, \, \bar\beta, \, \phi, \, B, \,{\cal B}, \,  B_{\mu},     
\bar B_{\mu}, \,C_\mu, \,  A_{\mu}, \, (\partial^\nu\, B_{\nu\mu}), 
  F_{\mu\nu} \Big] = 0, 
\end{eqnarray}
the coupled (but equivalent) Lagrangian densities ${\cal L}_B$ and  ${\cal L}_{\bar B}$ transform to the 
total spacetime derivatives as follows:
\begin{eqnarray}
&&s_{ad}\, {\cal L}_{B} = -\, \partial_\mu \, [{\cal B}\, \partial^\mu\, C - \rho \, \partial^\mu\, \beta] 
\equiv    s_{ad}\,{\cal L}_{\bar B}, \nonumber\\
&&s_d\, {\cal L}_{B} = -\, \partial_\mu \, [{\cal B}\, \partial^\mu\, \bar C + \lambda \, \partial^\mu\, \bar\beta] 
\equiv    
s_d \,  {\cal L}_{\bar B}. 
\end{eqnarray}
The above observations demonstrate that the action integrals, corresponding to the above Lagrangian densities ${\cal L}_B$ and  ${\cal L}_{\bar B}$,
would remain invariant under the (anti-)co-BRST symmetry transformations due to the validity of Gauss's divergence theorem
(where all the physical fields vanish off as $x \to \pm \infty$).

According to Noether's theorem, the invariance of the action integral under the continuous symmetry transformations {\it always}
leads to the derivations of the currents which are popularly known as the Noether currents. The conservation of these
currents is proven by using the EL-EoMs that are derived from the minimization of the action integral. In our present case, 
the above infinitesimal, continuous and off-shell nilpotent 
(anti-)dual BRST symmetry transformations [cf. Eqs. (60),(61)] lead to the following Noether currents:
\begin{eqnarray}
J^{\mu}_{(ad)} &=& \rho \,\partial^{\mu}  {\beta} -
\big (\partial^{\mu} \bar {C}^{\nu} -\partial^{\nu} \bar {C}^{\mu} \big ) \,\partial_{\nu} {\beta} 
- {\cal B} \, \partial^{\mu} C
- \varepsilon^{\mu \nu \sigma} \,\bar {B}_{\nu} \,\partial_{\sigma} C, \nonumber\\
J_{(d)}^{\mu} &=& - \, \Big [ {\mathcal{B}} \, \partial^{\mu} \bar {C} + \lambda \, \partial^{\mu} \bar {\beta}
+ \big (\partial^{\mu}C^{\nu} - \partial^{\nu} C^{\mu} \big ) \, \partial_{\nu} \bar {\beta}  
+ \varepsilon^{\mu\nu\sigma} \,B_{\nu}\, \partial_{\sigma} \bar {C} \Big ].
\end{eqnarray}
The conservation law (i.e. $\partial_\mu J^{\mu}_{(ad)} = 0 $) for the anti-co-BRST current $J^{\mu}_{(ad)} $ can be proven by
exploiting the beauty and strength of the following EL-EoMs
\begin{eqnarray}
&& \Box \,C =0 , \;\;\qquad  \Box \, {\beta} = 0, \;\;\qquad 
\partial_{\mu}(\partial^{\mu} \bar {C}^{\nu} - \partial^{\nu} \bar {C}^{\mu} ) - \partial^{\nu} \rho = 0, \nonumber\\
&&\varepsilon^{\mu \nu \sigma} \partial_{\sigma} {\cal B} + \big (\partial^{\mu}\bar{B}^{\nu} 
- \partial^{\nu} \bar {B}^{\mu} \big ) = 0
\quad \Longleftrightarrow \quad
\varepsilon^{\mu \nu \sigma} \,\partial_{\nu}\,\bar {B}_{\sigma} + \partial^{\mu} {\cal {B}} = 0,
\end{eqnarray}
which are derived from the Lagrangian density ${\cal L}_{\bar B}$. In exactly similar fashion, we observe
that the conservation law (i.e. $\partial_\mu J^{\mu}_{(d)} = 0 $) for the co-BRST current $J^\mu_{(d)}$ can be 
proven by using the theoretical strength of the following EL-EoMs 
\begin{eqnarray}
&& \Box \,\bar C =0 , \;\;\qquad  \Box \, \bar {\beta} = 0, \;\;\qquad 
\partial_{\mu}(\partial^{\mu}  {C}^{\nu} - \partial^{\nu}  {C}^{\mu} ) + \partial^{\nu} \lambda = 0, \nonumber\\
&&\varepsilon^{\mu \nu \sigma} \partial_{\sigma} {\cal B} + \big (\partial^{\mu} {B}^{\nu} 
- \partial^{\nu} {B}^{\mu} \big ) = 0
\quad \Longleftrightarrow \quad
\varepsilon^{\mu \nu \sigma} \,\partial_{\nu}\, {B}_{\sigma} + \partial^{\mu} {\cal {B}} = 0,
\end{eqnarray}
that are derived from the Lagrangian density ${\cal L}_B$. The conserved Noether (anti-)co-BRST currents (63) lead
to the definitions of the conserved charges as follows
\begin{eqnarray}
&&Q_{ad} = \int d^2 x\, J^0_{(ad)} \equiv +\, \int d^2 x\, \Big [\rho \, \dot \beta - {\cal B}\,\dot C 
- \big (\partial^0 \bar C^i - \partial^i \bar C^0 \big)\, \partial_i \beta - \varepsilon^{0ij} \,   B_i \, \partial_j C \Big ],
\nonumber\\
&& Q_{d} = \int d^2 x\, J^0_{(d)} \equiv -\, \int d^2 x\, \Big [\lambda \, \dot {\bar \beta} + {\cal B}\,\dot {\bar  C} 
+  \big (\partial^0  C^i - \partial^i  C^0 \big )\, \partial_i \bar \beta + \varepsilon^{0ij} \,   B_i \, \partial_j \bar C \Big ],
\end{eqnarray}
which are the generators for the continuous off-shell nilpotent (anti-)co-BRST symmetry transformations [cf. Eqs. (60),(61)] provided we use the
equation (22) with the replacements: $s_b \to s_{ad}, \; Q_b \to Q_{ad}$ and  $s_b \to s_{d}, \; Q_b \to Q_{d}$, respectively.

We end this section with the following remarks. First, we note that the gauge-fixing terms $(\partial \cdot A)$ and
$(\partial^\nu B_{\nu\mu} \pm \partial_\mu \phi)$ for the Abelian 1-form ($A^{(1)} = A_\mu\, dx^\mu $)
and 2-form [$B^{(2)} = \frac{1}{2!}\, B_{\mu\nu}\, (dx^\mu \wedge dx^\nu) $] gauge fields remain invariant, respectively,
under the (anti-) co-BRST symmetry transformations. Second, we would like to mention that the
nomenclature (anti-)co-BRST symmetries is {\it correct} because the above (anti-)co-BRST invariant 
gauge-fixing terms owe their origin to the co-exterior derivative (i.e. $\delta = \pm\, *\, d\, * $) of
differential geometry. To be precise, we note that $\delta  A^{(1)} = +\,* d\,* \, A^{(1)} = (\partial \cdot A) $ and
$\delta  B^{(2)} = -\,* d\,* \, B^{(2)} = (\partial^\nu B_{\nu\mu})\; d x^\mu$. In the gauge-fixing term 
$(\partial^\nu B_{\nu\mu} \pm \partial_\mu \phi)$
for the Abelian 2-form field, the scalar field $\phi$ (with proper mass dimension) appears due to 
the observation that there is existence of the stage-one reducibility
in the Abelian 2-form theory [29]. Third, in contrast to the (anti-)BRST symmetries, under which, the {\it individual} portions 
of the Lagrangian densities (16) and (36)  for the Abelian 1-form and 2-form theories remain invariant, under the (anti-)co-BRST
symmetry transformations {\it only} the FP-ghost term of the Abelian 1-form theory contributes along with the {\it rest} of the 
contributions coming from the Abelian 2-form theory for the derivation of (62). Thus, it is clear that for the existence of the
(anti-)co-BRST symmetries in our theory, the combined system of the free 
Abelian 1-form and 2-form theories should be taken together within the framework of  BRST formalism.
Fourth, exploiting the beauty and strength of the relationship between the 
infinitesimal and continuous symmetry
transformations and their generators as the Noether conserved charges, we obtain the following in the context 
of the infinitesimal, continuous, off-shell nilpotent (anti-)co-BRST symmetries and 
the  off-shell nilpotent (anti-)co-BRST charges:
\begin{eqnarray}
s_{ad}\, Q_{ad} &=& - i\, \{ Q_{ad}, \; Q_{ad} \} = 0 \quad \Longrightarrow \quad Q_{ad}^2 = 0, \nonumber\\
s_{d}\, Q_{d} &=& - i\, \{ Q_{d}, \; Q_{d} \} = 0 \qquad \Longrightarrow \quad Q_{d}^2 = 0.
\end{eqnarray}
In the above, the l.h.s. can be precisely computed by the {\it direct} application of the 
nilpotent (anti-)co-BRST symmetry transformations [cf. Eqs. (60),(61)]
on the explicit expression for the conserved and nilpotent (anti-)co-BRST charges (66). The equation (67) is nothing but the
proof that the (anti-)co-BRST charges are off-shell nilpotent (i.e. $Q_{(a)d}^2 = 0 $) of order two. Fifth, out
of the {\it four} existing nilpotent (anti-)BRST and (anti-)co-BRST transformations, the following anticommutativity relationships are satisfied, namely; 
\begin{eqnarray}
\{s_b, \, s_{ad}\} = 0, \qquad  \{s_b, \, s_{ab}\} = 0, \qquad  \{s_{ab}, \, s_{d}\} = 0, \qquad \{s_d, \, s_{ad}\} = 0,
\end{eqnarray}
where we need to invoke the validity of the CF-type restriction (i.e. $B_\mu - \bar B_\mu + 2\, \partial_\mu \phi = 0$) {\it only}
in the proof of the absolute anticommutaivity relationship between the BRST and anti-BRST symmetry operators:
$\{s_b, \, s_{ab}\} = 0$. Furthermore, we point out that the anticommuatativity property between the co-BRST and anti-co-BRST
symmetries are satisfied only up to the U(1) gauge transformations [29] as we observe
that the following are true
\begin{eqnarray}
\{s_d, \; s_{ad}\} \,\bar C_{\mu} = -\, \partial_\mu \rho, \qquad
\{s_d, \; s_{ad}\} \,C_{\mu} = -\, \partial_\mu \lambda, 
 \end{eqnarray}
 for the Lorentz vector (anti-)ghost fields $(\bar C_\mu)C_\mu$. It can be readily checked that the absolute anticommutativity
 property (i.e. $\{s_d, \; s_{ad}\} = 0 $) is satisfied for {\it all} the {\it other} fields of our theory. 
The left over anticommutator
relationships in (68) are automatically satisfied (i.e. $\{s_b, \, s_{ad}\} = 0,\; \{s_{ab}, \, s_{d}\} = 0 $) among the 
off-shell nilpotent transformation operators. Finally, the 
anticommutativity property  (i.e. $s_d \,s_{ad} + s_{ad} \,s_{d} $) of the 
off-shell nilpotent (i.e. $s_{(a)d}^2 = 0$)
(anti-)co-BRST symmetry transformations (up to the U(1) gauge transformations [cf. Eq. (69)]) is 
reflected in the proof of the anticommutativity property of the conserved and nilpotent (anti-)co-BRST charges (cf. Appendix C for details). \\


\section{Unique Bosonic Symmetry: Conserved Charge}

The purpose of this section is to derive the Noether conserved charge from a {\it unique} bosonic symmetry
transformation that emerges out from a {\it unique} anticommutator between the off-shell nilpotent (anti-)BRST
and (anti-)co-BRST symmetry transformations.  In this connection, it is pertinent to point out that, besides the 
{\it four} anticommutators that have been defined in (68), we have {\it two} more independent anticommutators which 
turn out to be non-zero and they define the following bosonic symmetry transformations:
\begin{eqnarray}
s_w = \{s_b, \, s_d \}, \qquad \qquad s_{\bar w} = \{ s_{ad}, \, s_{ab} \}. 
\end{eqnarray}
However, we have been able to demonstrate, in our earlier work [19], that $s_w + s_{\bar w} = 0 $. Hence, we have
a {\it unique} bosonic symmetry $s_w$ in our theory. Under this infinitesimal, local and continuous
bosonic symmetry transformation (i.e. $s_w = \{s_b, \, s_d \}$), the fields of our theory (described by the coupled 
Lagrangian densities ${\cal L}_B $ and ${\cal L}_{\bar B} $)  transform as\footnote{The Lorentz vector (anti-)ghost fields
$(\bar C_\mu)C_\mu$ transform as the U(1) gauge symmetry transformations just like our observations in (69) modulo a sign factor.
However, the transformations in (71) and (69) are in completely different contexts. In the case of the {\it latter}
where the proof of the anticommutator $\{ s_d, \; s_{ad} \} = 0$ has been discussed, {\it only} two fields 
$(\bar C_\mu)C_\mu$ have been found to transform  like the U(1) gauge symmetry transformations and rest of the fields
of our theory have respected the absolute anticommutativity property (i.e. $\{ s_d, \; s_{ad} \} = 0$). However, under
the bosonic symmetry transformations (71), the 2-form field $B_{\mu\nu}$ does {\it not} transform like the gauge symmetry
transformation even though the 1-form field $A_\mu$ does.}:
\begin{eqnarray}
&&s_w \, B_{\mu\nu}  = -\, \varepsilon_{\mu\nu\sigma}\, \partial^\sigma\, B, \qquad s_w \, \bar C_\mu = \partial_\mu\, \rho, \qquad
s_w\, C_\mu = \partial_\mu\, \lambda,  \qquad \nonumber\\ 
&& s_w\big [\rho, \, \lambda, \, \phi,  \, 
C,\, \bar C, \, \beta, \, \bar\beta, \,B, \, \, {\cal B}, \,B_\mu, \, \bar B_\mu, \,(\partial^\nu B_{\nu\mu}), \, F_{\mu\nu} \big ] = 0.
\end{eqnarray}
The above transformations are the {\it symmetry} transformations of our theory because we observe the following {\it equal} transformations
of ${\cal L}_B $ and ${\cal L}_{\bar B}$, namely;
\begin{eqnarray}
s_w \,{\cal L}_{B} = \partial_\mu\, \Big [B\, \partial^\mu\, {\cal B} - 
{\cal B} \, \partial^\mu\, B - \rho\, \partial^\mu\, \lambda\, + (\partial^\mu\, \rho)\, \lambda \Big] \equiv s_w \,{\cal L}_{\bar B},
\end{eqnarray}
which establishes that the action integrals, corresponding to the coupled (but equivalent) 
Lagrangian densities ${\cal L}_B $ and ${\cal L}_{\bar B} $, remain invariant under the infinitesimal, local and continuous 
bosonic symmetry transformation $s_w$ [cf. Eq. (71)].

According to Noether's theorem, the observation of the above invariance of the action integral [under the infinitesimal, local
and continuous bosonic symmetry transformations (71)] leads to the derivation of the Noether current $J^\mu_{(w)}$ as 
\begin{eqnarray}
J^\mu_{(w)} = F^{\mu\nu} \, \partial_\nu {\cal B} + \big (\partial^\mu \bar C^\nu - \partial^\nu \bar C^\mu \big )\, \partial_\nu \lambda
- \big (\partial^\mu  C^\nu - \partial^\nu  C^\mu \big )\, \partial_\nu \rho - \varepsilon^{\mu\nu\sigma} \, B_\nu \, \partial_\sigma B,
\end{eqnarray}
where the standard theoretical tricks of the Noether theorem have been exploited taking into account the observation in (72). The
conservation law (i.e. $\partial_\mu J^\mu_{(w)} = 0$) can be proven by making use of the following EL-EoMs
\begin{eqnarray}
&&  
\partial_{\mu}(\partial^{\mu}  \bar {C}^{\nu} - \partial^{\nu} \bar {C}^{\mu} ) - \partial^{\nu} \rho = 0, \qquad
\varepsilon^{\mu \nu \sigma} \,\partial_{\nu}\, {B}_{\sigma} + \partial^{\mu} {\cal {B}} = 0,
 \nonumber\\
&& \partial_\mu F^{\mu\nu} + \partial^\nu B = 0, \qquad \qquad
\partial_{\mu}(\partial^{\mu}  {C}^{\nu} - \partial^{\nu}  {C}^{\mu} ) + \partial^{\nu} \lambda = 0,
\end{eqnarray}
which are derived from the Lagrangian density ${\cal L}_B$. The conserved Noether current in (73) leads to the
definition of the conserved Noether charge $Q_w = \int d^2 x\, J^0_{(w)}$ as 
\begin{eqnarray}
Q_{w} = \int d^2 x\, \Big [ F^{0i} \, \partial_i {\cal B}  + \big (\partial^0  \bar C^i - \partial^i \bar C^0 \big )\, \partial_i \lambda
-  \big (\partial^0  C^i - \partial^i  C^0 \big )\, \partial_i \rho - \varepsilon^{0ij} \,   B_i \, \partial_j B \Big ]
\end{eqnarray}
which is the generator for the infinitesimal, local and continuous bosonic symmetry transformations (71) if we exploit the relationship (11)
with the replacements: $\delta g \to s_w, \; G \to Q_w$.

We conclude this section with the following useful and interesting remarks. First of all, we note that, due to the definition of 
$s_w = \{s_b, \; s_d \}$ and the off-shell nilpotency (i.e. $s_{(a)b}^2 = 0, s_{(a)d}^2 = 0 $) of  the (anti-)BRST
and (anti-)co-BRST symmetry transformations, the infinitesimal bosonic symmetry operator $s_w$ satisfies the following algebra: 
\begin{eqnarray}
[s_w , \; s_r ] = 0, \qquad r = b,\, ab,\, d,\, ad.
\end{eqnarray}
In other words, the symmetry operator $s_w$ commutes with {\it all} the off-shell nilpotent (i.e. $s_{(a)b}^2 = 0, s_{(a)d}^2 = 0 $)
(anti-)BRST and (anti-)co-BRST symmetry operators. Second, it is interesting to point out that we observe the following when
we apply the (anti-)co-BRST symmetry transformations {\it directly} on the expression for the bosonic charge
\begin{eqnarray}
s_{ad} Q_w = -\, i\, \big [Q_w, \; Q_{ad} \big ] = 0, \qquad s_d Q_w = -\, i\, \big [Q_w, \; Q_d \big ] = 0, 
\end{eqnarray}
which establishes that the bosonic charge $Q_w$ commutes with 
the conserved and nilpotent co-BRST and anti-co-BRST charges 
(i.e. $Q_d, \; Q_{ad} $) in a straightforward manner. Third, in exactly similar fashion, 
when we apply the BRST symmetry transformation (i.e. $s_b$) on the bosonic charge $Q_w$ [cf. Eq. (75)], we obtain the following explicit expression
\begin{eqnarray}
s_b Q_w = -\, i\, \big [Q_w, \; Q_b \big ] = - \,  \int d^2 x\, \big [ (\partial^0 B^i - \partial^i B^0)\, \partial_i \lambda \big ]
\equiv +\, \int d^2 x\, \partial_i \big [ (\partial^0 B^i - \partial^i B^0)\,  \lambda \big ], 
\end{eqnarray}
where we have applied the Gauss divergence theorem and dropped the total space derivative term. At this juncture, we apply the
following EL-EoM (on the r.h.s. of the final form)
\begin{eqnarray}
\varepsilon^{\mu \nu \sigma} \partial_{\sigma} {\cal B} + \big (\partial^{\mu} {B}^{\nu} 
- \partial^{\nu}  {B}^{\mu} \big ) = 0
\quad \Longleftrightarrow \quad \big (\partial^0 B^i - \partial^i B^0 \big ) = - \, \varepsilon^{0ij} \partial_{j} {\cal B},
\end{eqnarray}
which proves, in a straightforward manner, that we have: $\partial_i \big [ (\partial^0 B^i - \partial^i B^0) \big ] = 0$.
In other words, we find that the following is true, namely;
\begin{eqnarray}
s_b Q_w = - \, i\, \big [ Q_w, \; Q_b \big ] = 0  \qquad \Longrightarrow \qquad \big [ Q_b, \; Q_w \big ] = 0
\end{eqnarray}
which establishes the fact that, just like the (anti-)co-BRST charges [cf. Eq. (77)],  bosonic charge $Q_w$ commutes with 
the BRST charge, too. As we have performed our computations in the deduction of the commutation relation between $Q_w$ and $Q_b$,
the same logic and method can be applied to prove that the bosonic charge $Q_w$ commutes 
(i.e. $[Q_w, \; Q_{ab} ] = 0 $) with the non-nilpotent
but conserved anti-BRST charge $Q_{ab}$, too. These statements are valid for the non-nilpotent Noether conserved (anti-)BRST charges
as well as for the off-shell nilpotent versions of the (anti-)BRST charges $Q_{(A)B}$. To corroborate the {\it latter}
 claim, we can readily check that the following are true\footnote{It is straightforward to check that:
$s_w Q_B = \int d^2 x\, \big [(\partial^0 B^i - \partial^i B^0)\, \partial_i \lambda \big ]$. Using the Gauss divergence theorem and 
the EL-EoM: $\varepsilon^{0ij} \,\partial_j {\cal B} + (\partial^0 B^i - \partial^i B^0) = 0 $, this expression can be recast
in the following form: $ s_w Q_B = \int d^2 x\,\varepsilon^{0ij}\, (\partial_i \,\partial_j {\cal B})\, \lambda$.
This observation  automatically implies that the expression $s_w Q_B $ is equal to {\it zero} which, in turn, leads to 
the derivation of the commutator: $[Q_B. \; Q_w ] = 0 $. In exactly similar manner, one  can exploit  the (i) Gauss divergence theorem, and (ii) 
appropriate EL-EoM to prove that the nilpotent anti-BRST charge $Q_{AB} $ {\it also} commutes with $Q_w$ (i.e. $[Q_{AB}. \; Q_w ] = 0 $).}, namely;
\begin{eqnarray}
s_{w} Q_B = -\, i\, \big [Q_B, \; Q_{w} \big ] = 0, \qquad s_w Q_{AB} = -\, i\, \big [Q_{AB}, \; Q_w \big ] = 0, 
\end{eqnarray}
where the l.h.s. can be computed explicitly by applying the bosonic symmetry transformations (71) on the explicit 
expressions in (57) and (32) which are nothing but the {\it nilpotent} versions of the (anti-)BRST charges $Q_{(A)B}$, respectively.
Finally, we have the following form of the algebraic relationship between the bosonic charge $Q_w$ and the off-shell nilpotent versions of the
(anti-)co-BRST and (anti-)BRST charges [cf. Eqs.(57),(32)], namely;
\begin{eqnarray}
\big [ Q_w, \; Q_r \big ] = 0, \qquad r =  d,\, ad, \, B,\, AB.
\end{eqnarray}
The above observation establishes that the bosonic charge commutes with {\it all} the off-shell nilpotent 
versions of the conserved charges of our theory.\\


\section{Ghost-Scale Transformations: Ghost Charge}

It is very interesting to point out that under the following ghost-scale transformations 
\begin{eqnarray}
&&\beta \longrightarrow e^{+ 2\, \Sigma} \, \beta,  \qquad \quad
\bar \beta \longrightarrow e^{-\,  2\, \Sigma} \,\bar\beta, \quad
C_\mu \longrightarrow e^{+ \Sigma} \, C_\mu,  \qquad \quad
\bar C_\mu \longrightarrow e^{-\, \Sigma} \,  \bar C_\mu, \nonumber\\
&& C \longrightarrow e^{+ \Sigma} \,  C, \qquad \qquad \; \;
\bar C \longrightarrow e^{-\,  \Sigma} \,  \bar C, \qquad
\lambda \longrightarrow e^{ + \Sigma} \,  \lambda, \qquad 
\rho \longrightarrow e^{-\, \Sigma} \,  \rho, \nonumber\\
&&\Phi \longrightarrow e^0\; \Phi \qquad \qquad (\Phi = 
A_\mu, \, B_{\mu\nu}, \, F_{\mu\nu},\,  \bar B_\mu, \, B_\mu, \, \phi, \, {\cal B}, \,B),  
\end{eqnarray}
where $\Sigma$ is a global (i.e. spacetime independent) scale transformation parameter, the FP-ghost part of the
Lagrangian densities
${\cal L}_{B}$ and ${\cal L}_{\bar B}$ remain non-trivially invariant because the {\it rest} part of
{\it these} Lagrangian densities are trivially invariant due to the generic transformation: $\Phi \longrightarrow e^0\; \Phi $ in (83).
It is worthwhile to point out that the numerals, in front of the scale transformation parameter $\Sigma$ in the exponents,
are  nothing but the appropriate ghost numbers for the (anti-)ghost fields.  For the sake of brevity, 
if we choose the scale parameter $\Sigma = 1$, the infinitesimal version  of the {\it above} ghost-scale 
symmetry transformations $(s_g)$ reduces to the following transformations of the fields of our theory, namely;
\begin{eqnarray}
&&s_g\, \beta = + 2\, \beta, \qquad s_g\, \bar \beta = -\, 2\, \bar\beta, \qquad 
s_g\, C_\mu = +\, C_\mu, \qquad
s_g\, \bar C_\mu = -\, \bar C_\mu,  
 \nonumber\\
&& s_g\, C = + \, C, \qquad s_g\, \bar C = -\, \bar C, \qquad
s_g\, \lambda =  \lambda, \qquad 
s_g\, \rho = -\, \rho, \quad s_g\, \Phi = 0.
\end{eqnarray}
It is straightforward to check that, under the above infinitesimal version of the ghost-scale symmetry transformations, we obtain: 
$s_g\, {\cal L}_{B} = 0, \; s_g\, {\cal L}_{\bar B} = 0$. Hence, the action integrals, corresponding to the 
Lagrangian densities ${\cal L}_{B}$ and ${\cal L}_{\bar B}$, 
{\it also} remain invariant under the infinitesimal version of the 
bosonic (i.e. $s_g^2 \neq 0 $) ghost-scale symmetry transformation $s_g$.

The above observation of the invariance of the Lagrangian densities, according to the celebrated Noether theorem, leads to the 
definition of the conserved ghost current $J^\mu_{(g)}$ as:
\begin{eqnarray}
J^\mu_{(g)} &=& 2\, \beta \, \partial^\mu \bar \beta - 2\, \bar \beta \, \partial^\mu  \beta + \lambda\,  \bar C^\mu-
\rho\,   C^\mu + \bar C \, \partial^\mu C -  (\partial^\mu \bar C) \, C
\nonumber\\
&+& \big (\partial^\mu C^\nu - \partial^\nu C^\mu \big )\,
\bar C_\nu + \big (\partial^\mu \bar C^\nu - \partial^\nu \bar C^\mu \big )\, C_\nu. 
\end{eqnarray}
The conservation law ($\partial_\mu J^\mu_{(g)} = 0 $) of the above current can be proven by using the following EL-EoMs
that emerge out from {\it both} the Lagrangian densities  ${\cal L}_{B}$ and ${\cal L}_{\bar B}$, namely;
\begin{eqnarray}
&& \Box \,C =0, \quad  \Box \, {\beta} = 0, \quad 
\partial_{\mu}(\partial^{\mu} \bar {C}^{\nu} - \partial^{\nu} \bar {C}^{\mu} ) - \partial^{\nu} \rho = 0, 
\quad  \Box \,\bar C =0, \quad  \Box \, \bar {\beta} = 0, \nonumber\\
&& \partial_{\mu}(\partial^{\mu}  {C}^{\nu} - \partial^{\nu} {C}^{\mu} ) + \partial^{\nu} \lambda = 0, 
\qquad \lambda = \frac{1}{2}\, \big (\partial \cdot C \big ), \qquad  \rho = -\, \frac{1}{2}\, \big (\partial \cdot \bar C \big ).
\end{eqnarray}
The conserved current in (85) leads to the definition of the conserved ghost charge $Q_g$  as:
\begin{eqnarray}
Q_g = \int d^2 x\, J^0_{(g)} &\equiv & \int d^2 x\, \Big [ 2\, \beta \, \dot {\bar \beta} - 2\, \bar \beta \, \dot  \beta + \lambda\,  \bar C^0 -
\rho\,   C^0 + \bar C \, \dot C -  \dot {\bar C} \, C 
\nonumber\\
&+& \big (\partial^0 C^i - \partial^i C^0 \big )\,
\bar C_i + \big (\partial^0 \bar C^i - \partial^i \bar C^0 \big )\, C_i \Big ]. 
\end{eqnarray}
The above charge turns out to be  the generator for the infinitesimal version of the ghost-scale symmetry transformations (84) provided we use the
theoretical strength of the relationship in (11) with the replacements: $\delta_g \to s_g, \; G \to Q_g$.

We conclude this section with the following remarks. First of all, we note that, under the ghost-scale symmetry transformations, {\it only}
the (anti-)ghost fields transform non-trivially with a global (i.e. spacetime-independent) scale transformation parameter in the exponent. The
numerals in front of {\it it} correspond to the ghost numbers. The non-ghost (i.e. the physical) fields 
of our theory do {\it not} transform at all under
the ghost-scale symmetry transformations [cf. Eqs. (83),(84)].  Second, we observe that, in addition to the continuous symmetry
transformations  (84), the FP-ghost part of the 
coupled (but equivalent) Lagrangian densities  ${\cal L}_{B}$ and ${\cal L}_{\bar B}$ respects the following discrete symmetry transformations:
\begin{eqnarray}
&&C_\mu \to \pm\, i\, \bar C_\mu, \qquad \bar C_\mu \to \pm\, i\,  C_\mu, \qquad C \to \pm\, i\, \bar C, \qquad \bar C \to \pm\, i\, C,
\nonumber\\
&&\beta \to \pm\, i\, \bar \beta, \qquad \bar \beta \to \mp\, i\, \beta, \qquad \lambda \to \mp i\, \rho, \qquad \rho \to \mp i\, \lambda.
\end{eqnarray}
Third, we observe that the ghost charge $Q_g$ obeys the following algebra with the {\it rest} of the fermionic
(i.e. off-shell nilpotent) and bosonic conserved charges of the theory, namely;
\begin{eqnarray}
&& s_g Q_b = - \, i\, \big [ Q_b, \; Q_g \big ] = + \, Q_b  \qquad \quad \Longrightarrow  \qquad i\, \big [ Q_g, \; Q_b \big ] = + \; Q_b, \nonumber\\
&& s_g Q_{ab} = - \, i\, \big [ Q_{ab}, \; Q_g \big ] = - \, Q_{ab} \qquad \Longrightarrow \qquad i\, \big [ Q_g, \; Q_{ab} \big ] = - \; Q_{ab}, \nonumber\\
&& s_g Q_d = - \, i\, \big [ Q_d, \; Q_g \big ] = - \, Q_d \qquad \quad \Longrightarrow  \qquad i\, \big [ Q_g, \; Q_d \big ] = - \; Q_d, \nonumber\\
&& s_g Q_{ad} = - \, i\, \big [ Q_{ad}, \; Q_g \big ] = + \, Q_{ad} \qquad \Longrightarrow \qquad i\, \big [ Q_g, \; Q_{ad} \big ] = + \; Q_{ad}, \nonumber\\
&& s_g Q_w = - \, i\, \big [ Q_w, \; Q_g 
\big ] = 0 \qquad \qquad \;\;\Longrightarrow \qquad  i\, \big [ Q_g, \; Q_w \big ] = 0. 
\end{eqnarray}
It is trivial to point out that $s_g Q_g = - i\, [Q_g, \; Q_g ] = 0$. Fourth, we would like to lay emphasis on the fact
 that the {\it unique} bosonic charge $Q_w$ commutes with all the {\it rest} of the charges of our theory as is clear from 
our observations in the last entry of the above equation [cf. Eq. (89)] and (82).
Finally, the above algebra (with the ghost conserved charge) implies that if we define
the ghost number of a quantum state $|\psi>_n$ as: $i\, Q_g \, |\psi>_n = n\, |\psi>_n $, the following relationships emerge out
very naturally, namely; 
\begin{eqnarray}
i \, Q_g \, Q_b \, |\psi>_n & = & (n + 1) \, Q_b \, |\psi>_n, \qquad  i \, Q_g \, Q_d \, |\psi>_n  =  (n - 1) \, Q_d \, |\psi>_n, \nonumber\\
i \, Q_g \, Q_{ad} \, |\psi>_n & = & (n + 1) \, Q_{ad} \, |\psi>_n, \quad \; i \, Q_g \, Q_{ab} \, |\psi>_n  =  (n - 1) \, Q_{ab} \, |\psi>_n, \nonumber\\
i \, Q_g \, Q_w \, |\psi>_n & = & (n + 0) \, Q_w \, |\psi>_n. 
\end{eqnarray}
The above relationships demonstrate that there are two quantum states $ Q_b \, |\psi>_n $ and $ Q_{ad} \, |\psi>_n $ which possess
the ghost number $(n + 1)$. On the other hand, we have two states $ Q_d \, |\psi>_n $ and $ Q_{ab} \, |\psi>_n $
in our theory that are endowed with the ghost number $(n - 1)$. In other words, the operation of the pair of 
conserved and nilpotent charges ($Q_b, \; Q_{ad} $)
on a quantum state raises the ghost number by one. On the contrary, the ghost number of a quantum state is lowered by one when it is
operated upon by the pair of conserved and nilpotent charges $(Q_d, \; Q_{ab})$. It interesting to point out that the ghost number 
of a quantum state remains intact when
it is operated upon by the {\it unique} bosonic charge $Q_w$. These observations\footnote{When the exterior derivative acts on a given form (i.e. $f_n$)
of degree $n$, it raises the degree of the ensuring form by one (i.e. $d \,f_n \to f_{n + 1}$). On the other hand, the degree of a form 
is lowered by one when it is operated upon by the co-exterior derivative (i.e. $ \delta\, f_n \to f_{n - 1}$). The degree of a form remains
unchanged when the Laplacian operator $\Delta = (d + \delta)^2 \equiv \{ d, \; \delta \} $  acts on it (i.e. $\Delta \,f_n \to f_{n}$).   }
are analogous to the operations of the de Rham cohomological operators of the differential geometry on a given form of degree $n$. \\

\section{Extended BRST Algebra: Appropriate Conserved Charges and Cohomological Operators}

The standard BRST algebra is satisfied among the 
conserved and off-shell nilpotent  (i.e. $Q_{(A)B}^2 = 0$) versions of the (anti-)BRST charges $Q_{(A)B}$
and the conserved ghost charge $Q_g$ as: $ Q_{(A)B}^2 = 0, \; i \, [Q_g, \; Q_B ] = +\, Q_B, \; i \, [Q_g, \; Q_{AB} ] = -\, Q_{AB}$.
However, in our present field-theoretic example for Hodge theory, we have total {\it six} appropriate conserved charges. These 
appropriately chosen charges
obey the following {\it extended} BRST algebra:
\begin{eqnarray}
&& Q_{(A)B}^2 = 0, \qquad Q_{(a)d}^2 = 0, \qquad  \{Q_B, \; Q_{ad} \} = 0, \qquad
 \{Q_d, \; Q_{AB} \}= 0, \quad  i\, [Q_g, \; Q_g ] = 0, \nonumber\\
&&   i \, [Q_g, \; Q_B ] = + Q_B, \quad i \, [Q_g, \; Q_{AB} ] = - Q_{AB}, \quad
  i \, [Q_g, \; Q_{ad} ] = + Q_{ad}, \quad  i \, [Q_g, \; Q_{d} ] = - Q_{d}, \nonumber\\
&&   Q_w = \{Q_B, \; Q_{d} \} \equiv  -\, \{Q_{AB}, \; Q_{ad} \}, \qquad [Q_w, \; Q_r] = 0 \quad (r = B, AB, d, ad, g, w).  
\end{eqnarray}
A few decisive comments, at this juncture, are in order. First of all, we observe that the commutators: $i \, [Q_g, \; Q_B ] = +\, Q_B $ and
$i \, [Q_g, \; Q_b ] = +\, Q_b$ are equivalent as are  the commutators: $i \, [Q_g, \; Q_{AB} ] = -\, Q_{AB} $
 and $i \, [Q_g, \; Q_{ab} ] = -\, Q_{ab} $ where the Noether conserved charges $Q_{(a)b}$ are non-nilpotent (i.e. $Q^2_{(a)b} \ne 0$). 
Second, the anticommutativity property between the nilpotent BRST and anti-BRST charges is
 valid only if we invoke the validity of the CF-type restriction: $B_\mu - \bar B_\mu + 2\, \partial_\mu \phi = 0$ 
which is the reflection of the observation made in (44) where we observe that $\{ s_b, \; s_{ab} \}\, B_{\mu\nu} = 0 $ if and only if 
the CF-type restriction is invoked in the physical-sector where the gauge field $B_{\mu\nu}$ is defined. On the other hand,
in the ghost-sector, we note that the anticommuativity is valid modulo the U(1) gauge symmetry-type transformations [cf. Eq, (44)].
Third, the observations in (44) are reflected in the requirement of the anticommutatvity property between the conserved and
nilpotent versions of the (anti-)BRST charges (cf. Appendix B for details). 
Fourth, the anticommutativity property between the co-BRST and anti-co-BRST symmetries is {\it true} only modulo
 a U(1) gauge symmetry-type transformations [cf. Eq. (69)]. This observation is reflected in the proof of the
 anticommutativity property between the conserved and nilpotent co-BRST  and anti-co-BRST charges (cf. Appendix C). Fifth, as far as the
 validity of the anticommutators: $\{Q_B, \; Q_{ad} \} = 0, \;\{Q_d, \; Q_{AB} \}= 0 $ are concerned, we note that:
 $s_{ad} Q_B = -i\, \{Q_B, \; Q_{ad} \} \equiv - \, \int  d^2 x\, \big (\partial^0 B^i - \partial^i B^0 \big )\, \partial_i \beta$ and
  $s_{d} Q_{AB} = -i\, \{Q_{AB}, \; Q_{d} \} \equiv  - \,\int  d^2 x\, \big (\partial^0 \bar B^i - \partial^i \bar B^0 \big )\,  \partial_i\bar \beta$.
  Using the Gauss divergence theorem and the appropriate EL-EoMs [cf. Eqs. (79),(64)], it can be readily shown that: $s_{ad} Q_B   \equiv
\int \big (\varepsilon^{0ij}\, \partial_i \partial_j {\cal B} \big )\, \beta$ and $ s_{d} Q_{AB}= \int \big (\varepsilon^{0ij}\, \partial_i \partial_j {\cal B} \big )\, \bar \beta $ are equal to {\it zero}. 
Sixth, the bosonic charge $Q_w$ behaves like the Casimir operator for the whole extended BRST algebra. However, it is {\it not} in the sense 
 of the Casimir operators that are defined in the  contexts of the Lie algebras. Finally, we derive explicitly the 
{\it unique} bosonic charge $Q_w$ from
 the anticommutators $ \{Q_B, \; Q_{d} \} $ and $-\, \{Q_{AB}, \; Q_{ad} \}$ in our Appendix D.

 A close look at the extended BRST algebra (91) shows that it is reminiscent of the Hodge algebra that is obeyed by the
 de Rham cohomological operators ($d, \, \delta,\, \Delta $) of differential geometry\footnote{On a compact spactime manifold
without a boundary, the set of three operators ($d, \delta, \Delta$) is called as  the 
set of de Rham cohomological operators of differential geometry where 
$d$ (with $d^ 2 = 0$) is the exterior derivative, $\delta = \pm *\, d\, *$ (with $\delta^2 = 0 $) is the co-exterior derivative and
$\Delta = (d + \delta)^2 \equiv \{d, \; \delta \} $ is the Laplacian operator. Here $*$ is the Hodge duality operator (that is defined
on the {\it above} compact spactime manifold). These operators obey an algebra [cf. Eq. (92) below] which is popularly known as the Hodge algebra.}. 
The {\it latter}  algebra is as  follows [15-18] 
\begin{eqnarray}
&&d^2 = 0 \qquad \delta^2 = 0, \qquad \Delta = (d + \delta)^2 = \{ d, \; \delta \}, \nonumber\\
&& [\Delta, \; d] = 0, \qquad [\Delta, \; \delta] = 0, \qquad \{ d, \; \delta \} \neq 0,
\end{eqnarray}
where $\Delta = (d + \delta)^2 $ is the Laplacian operator which is always positive definite (and it behaves like the 
Casimir operator for the whole algebra but {\it not} in the Lie algebraic sense). A comparison between (91) and (92) establishes that there exists a 
two-to-one mapping between the conserved charges of our theory and the cohomological operators, namely;
\begin{eqnarray}
\big (Q_B, \,Q_{ad} \big) \longrightarrow d, \qquad \big (Q_d, \,Q_{AB} \big) \longrightarrow \delta, \qquad
\big \{ Q_B, \,Q_{d} \big\}  = Q_w \equiv  -\, \big \{ Q_{AB}, \,Q_{ad} \big \} \longrightarrow \Delta.
\end{eqnarray}
We would like to point out that if {\it only} the (anti-)BRST symmetries exist in a theory, we can {\it not} identify the {\it corresponding} 
BRST charge with the exterior derivative ($d$) and the anti-BRST charge with the co-exterior derivative ($\delta$) because
they [i.e. the conserved and nilpotent (anti-)BRST charges] anticommute with each-other (if we invoke the CF-type restriction).
However, it is clear from the Hodge algebra (92) that the exterior derivative and co-exterior derivative do {\it not}
anticommute with each-other. We lay emphasis on the fact that, only after the definitions and derivations
 of the (anti-)co-BRST charges $Q_{(a)d}$, we can be able to show the kind of two-to-one mapping that is illustrated in (93).

We wrap-up this section with the following remarks. First, our 3D field-theoretic system is an example for the Hodge theory only
at the {\it algebraic} level. Second, we have {\it not} been able to find out a set of discrete duality symmetry transformations
in our present theory (unlike our earlier works [36, 37, 6-9] in the cases of the 2D and 4D field-theoretic models) which provide the
physical realization(s) of the Hodge duality $*$ operator of differential geometry  in the relationship: $\delta = \pm\, *\, d\, * $.
Third, as is clear from our observations in (44), we have the absolute anticommutativity of the BRST and anti-BRST symmetries (i)
due to the validity of the CF-type restriction in the physical-sector where the gauge field $B_{\mu\nu}$ is defined, and (ii) modulo
the Abelian U(1) gauge symmetry-type transformations in the ghost-sector. These observations have been found to be reflected in the
requirement of the absolute anticommutativty between the conserved and nilpotent BRST and anti-BRST charges (cf. Appendix B).
Fourth, we have seen that the (anti-)co-BRST symmetries and corresponding charges do {\it not} absolutely anticommute,
Rather, their antcommutativity property is valid {\it only} modulo the  U(1) gauge symmetry-type transformations [cf. Eq. (69) and
Appendix C].  In our earlier works [36, 37, 6-9] on the 2D and 4D theories, we have been able to find out the {\it absolutely} anticommuting
(anti-)co-BRST symmetries (and corresponding charges). Finally, against the backdrop of all these lacunae, we might be tempted
to call our present 3D field-theoretic system as an example for a {\it quasi}-Hodge theory.\\

\section{Conclusions}

In our present investigation, we have performed a thorough constraint analysis for the D-dimensional combined system of the
free Abelian 1-form anf 2-form gauge theories and established that these constraints are of the first-class variety 
in the terminology of Dirac's prescription for the classification scheme of constraints [21-27]. We have commented on the special features
that are associated with the 3D combined system of the Abelian 1-form and 2-form gauge theories [cf. Eqs. (13),(14),(15)] and generalized
the {\it classical} gauge symmetry transformations (6) to their 
{\it quantum} counterparts nilpotent BRST and anti-BRST symmetry transformations [cf. Eqs. (17),(37)]. However, these {\it latter}
nilpotent symmetries do {\it not} lead to the derivation of the nilpotent {\it Noether} conserved charges. To circumvent this issue, 
we have derived the off-shell 
nilpotent version of the (anti-)BRST charges [cf. Eqs. (57),(32)] from the non-nilpotent versions of the {\it Noether} 
(anti-)BRST charges [cf. Eqs. (41),(21)]. We have also discussed the physicality criteria w.r.t. the off-shell nilpotent versions
of the (anti-)BRST charges and shown that the operator forms of the first-class constraints (of the
{\it classical}  combined system of the Abelian 1-form and 2-form gauge theories) annihilate the physical states at the
{\it quantum} level [cf. Eqs. (59),(34)] which are found to be consistent with the famous Dirac's quantization conditions
for the physical systems that are endowed with constraints (see, e.g. [26]).

We would like to add a few more sentences in connection with the non-nilpotent Noether (anti-)BRST charges $Q_{(a)b}$ and their
nilpotent versions $Q_{(A)B}$. Both the sets of charges can be exchanged with each-other without spoiling the conservation law if we
use at appropriate places (i) the Gauss divergence  theorem, (ii) the EL-EoMs from the Lagrangian densities (36) and (16). 
To make this point clearer, let us focus on the Noether BRST charge $Q_b$ [cf. Eq. (21)] and its nilpotent version $Q_B$ [cf. Eq. (32)].
Since both the charges are {\it equivalent}, we observe that
both can be used (i) as the generators of the BRST symmetry transformations (17), and (ii) in the
physicality criterion w.r.t. the BRST charge. However, we hasten to add that the Noether conserved charge $Q_b$ is {\it easier} to handle
in the relationship (22) as the generator for the BRST transformations (17) in contrast to the nilpotent version of the BRST charge 
$Q_B$ [cf. Eq. (32)]. In the {\it latter} case, one has to be more careful to obtain the {\it correct} BRST transformations (17). In exactly similar fashion, the nilpotent version of the BRST charge $Q_B$ is used to produce the quantization conditions (34) on the physical states which implies that the
operator forms  of the {\it primary} and {\it secondary} constraints annihilate the physical states {\it together}. On the other hand, 
the Noether conserved charge $Q_b$, in the physicality criterion w.r.t. the BRST charge, leads to the annihilation of
physical states  {\it only} by the primary constraints in a straightforward manner. Hence, if one uses the Noether BRST charge 
in the physicality criterion, one has to be more careful in obtaining the {\it correct} quantization conditions which require that the operator form of
{\it all} the constraints (i.e. primary, secondary, etc.) must annihilate the physical states {\it together}.

A few highlights of our present endeavor are as follows. First of all, we note that the absolute anticommutativity  (i.e. $\{s_b, \; s_{ab}\} = 0 $)
property between the BRST and anti-BRRT symmetry transformations is found to be {\it true} if and only if (i) the CF-type restriction 
(i.e. $B_\mu - \bar B_\mu + 2\,\partial_\mu \phi = 0 $) is invoked 
in the physical-sector where the gauge field $B_{\mu\nu}$ is defined [cf. Eq. (44)], and
(ii) modulo the Abelian U(1) gauge symmetry-type transformations in the ghost-sector 
which is not a {\it new} observation (see, e.g. [29],[38] for details).
Second, the absolute anticommutativity  (i.e. $\{s_d, \; s_{ad}\} = 0 $)
of the co-BRST and anti-co-BRST symmetry transformations is valid {\it only} modulo the Abelian U(1) gauge symmetry-type transformations [cf. Eq. (69)
and the corresponding footnote]. Third, we observe that, under the
{\it unique} bosonic transformations [cf. Eq. (71)], the FP-ghost fields {\it either} do not transform at all  {\it or} 
they transform up to the Abelian U(1) gauge symmetry-type transformations. Fourth, under the ghost-scale transformations [cf. Eqs. (83),(84)],
we note that {\it only} the FP-ghost fields transform and the {\it rest} of the fields  do {\it not} transform at all. Fifth, 
the kinetic terms (owing their origin to the exterior derivative of differential geometry) 
of the Abelian 1-form and 2-form gauge theories remain invariant under the (anti-)BRST 
transformations. Sixth, the gauge-fixing terms (owing their origin to the co-exterior derivative of differential geometry) are
found to be invariant under the nilpotent (anti-)co-BRST  transformations. Seventh, for our present 3D field-theoretic system, it turns out
that the kinetic as well as the gauge-fixing terms of the Abelian 1-form gauge field remain invariant under the (anti-)co-BRST
transformations. However, for the free Abelian 2-form field, the kinetic term changes under the (anti-)co-BRST
transformations {\it but} the gauge-fixing term remains invariant. Finally, under the (anti-)BRST transformations,
the Lagrangian densities of the Abelian 1-form and 2-form theories remain invariant separately and independently. However, under the off-shell
nilpotent (anti-)co-BRST transformations, {\it only} the FP-ghost fields of the Abalein 1-form gauge theory transform and the {\it rest} of the 
transformations are connected with the fields of the 3D Abelian 2-form gauge theory. Hence, for the existence of the off-shell nilpotent
(anti-)BRST and (anti-)co-BRST symmetries {\it together} in our 3D BRST-quantized theory, it is essential that the combined system of 
the free Abelian 1-from and 2-form gauge theories should be considered together.

It is worthwhile to point out that, unlike the 2D and 4D field-theoretic examples for Hodge theory [6-8,36,37], our present 3D example of 
Hodge theory does not support the existence of the fields with negative kinetic terms. To be precise, we have been able to show, in our
earlier works on the above 2D and 4D models of Hodge theory, the existence of a pseudo-scalar field and an axial-vector field with the
(i)  negative kinetic terms, and (ii)  well-defined rest masses. They have appeared in the above theories because of  the
symmetry considerations {\it alone}. Such fields have become quite popular in the realm of the cyclic, bouncing and self-accelerated cosmological
models of the Universe (see, e.g. [39-41] and references therein) where these fields  have been christened as the ``phantom'' and/or ``ghost'' fields.
Furthermore, such kinds of fields have been treated as the possible candidates of dark matter (see, e.g. [42,43] and references therein)
because of their ``exotic'' properties. In our
present 3D combined system of the Abelian 1-form and 2-form gauge theories (as the field-theoretic example for Hodge theory), there  
is no room for the existence of such kinds of fields with the negative kinetic terms. Even in our recent work on the St$\ddot {u}$ckelberg-modified
massive 3D Abelian 2-form (BRST-quantized) theory [32], there is no existence of the pseudo-scalar and/or axial-vector field with the negative kinetic terms.
Perhaps, this is the special feature of the {\it odd} dimensional (i.e. D = 3) field-theoretic example for Hodge theory that the 
fields with negative kinetic terms do {\it not} exist.

As pointed out at the fag end of Sec. 8, there are some {\it new} features associated with our 3D combined system of the Abelian 1-form
and 2-form gauge theories which turn out to be a model of Hodge theory. For instance, it appears to us that there is {\it no} 
existence of the discrete duality symmetry 
transformations in our theory because we do {\it not} have the existence of the pseudo-scalar field ($\tilde \phi$)
which is {\it dual} to the scalar field ($\phi$) of our theory. The {\it former}
 exists in the case of the 2D models of massless as well as St$\ddot {u}$ckelberg-modified
massive Abelian theories (see, e.g. [20,36,37]) which are the examples for Hodge theory. It is pertinent to point out that, even though
we have the existence of the discrete symmetries (85) in the ghost-sector, there is {\it no}  possibility of having the discrete
symmetry transformations in the physical-sector
because of the {\it absence} of the pseudo-scalar field. Thus, there is no analogue of the Hodge duality $*$ operator in our theory. Another issue 
which is different from our earlier works [6-8,20,36,37] is the observation that (i) the co-BRST and anti-co-BRST symmetry transformations
do not absolutely anticommute with each-other. Rather, their anticommutativity is valid only up to the U(1) gauge symmetry-type
transformations  in the ghost-sector [cf. Eq. (69)], and (ii) the anicommutativity of the BRST
and anti-BRST symmetries is also valid modulo the Abelian U(1) gauge symmetry-type transformations in the ghost-sector [cf. Eq. (44)].  
Perhaps, one has to introduce {\it another} CF-type restriction to resolve these issues
as is the case in the 4D Abelian 2-form and 6D Abelian 3-form (see, e.g. [6-8]) BRST-quantized theories. These are the issues we 
plan to address in our future endeavor  on the BRST approach [11-14] to the 3D field-theoretic example for Hodge theory 
which is a combined system of the free Abelain 1-form and 2-form gauge theories. The St$\ddot {u} $ckelberg-modified massive
version of the above 3D system is yet another direction which we would like to pursue in our future investigation [44].\\

\vskip 0.5cm

\noindent
{\bf Acknowledgments}\\

\noindent
The idea behind our present work originated when one of us (RPM) visited the School of Physics, University of Hyderabad (UoH). 
He deeply appreciates the  IoE-UoH- IPDF (EH) scheme of  UoH for taking care of his  travel  and local hospitality. Two of us
(B.R. and H.S.) express their deep sense of gratitude towards the DST-INSPIRE program  (IF220179) 
and the Prime Minister Research Fellowship (PMRF id:3703690), respectively, for the 
financial support under which the present investigation has been carried out. All the authors thank the Reviewer for very useful
comments which have certainly made the presentation of this paper more readable, lucid and transparent. \\

\vskip 0.3cm

\noindent
{\bf Data Availability:} No data were used to support this study.\\

\vskip 0.3cm

\noindent
{\bf Conflicts of Interest:}
The authors declare that there are no conflicts of interest. \\

\vskip 0.5cm

\begin{center}
{\bf Appendix A: On the Equivalence of ${\cal L}_B$ and ${\cal L}_{\bar B}$ and 
the Derivation of the (Anti-)BRST and (Anti-)-co-BRST Invariant CF-Type Restriction}\\
\end{center}

\vskip 0.3cm

\noindent
As far as the points of view of symmetries are concerned, we have been able to show the equivalence of the 
Lagrangian densities ${\cal L}_B$ and ${\cal L}_{\bar B}$ w.r.t. the (anti-)BRST symmetry 
transformations [cf. Eqs. (18),(38),(46),(47)] as well as the (anti-)co-BRST symmetry transformations 
[cf. Eq. (62)] where we have been able to show that {\it both} the Lagrangian densities  ${\cal L}_B$ and ${\cal L}_{\bar B}$ 
respect {\it all} the above {\it four} fermionic (i.e. off-shell nilpotent) symmetry transformations provided 
we invoke the validity of the CF-type restriction: $B_\mu - \bar B_\mu + 2\, \partial_\mu \phi = 0$ [cf. Eqs. (46),(47)].
In our present Appendix, we show the existence of the above CF-type restriction by demanding the {\it direct} equality of the 
Lagrangian densities ${\cal L}_B$ and ${\cal L}_{\bar B}$ (modulo a total spacetime derivative term). In other words, we
wish to establish that $ {\cal L}_{\bar B} - {\cal L}_B = 0$ if and only if the CF-type restriction: $B_\mu - \bar B_\mu + 2\, \partial_\mu \phi = 0$
is satisfied. A close look at the coupled Lagrangian densities  ${\cal L}_B$ and ${\cal L}_{\bar B}$ [cf. Eqs. (16),(36)]
shows that we have the following explicit difference between them:
\[
{\cal L}_{\bar B} - {\cal L}_B = \frac{1}{2} \, B^\mu\, B_\mu  - \frac{1}{2} \, \bar B^\mu\, \bar B_\mu
+ \bar B^\mu \,\big(\partial^\nu B_{\nu\mu} + \partial_\mu \phi \big ) - 
 B^\mu \,\big(\partial^\nu B_{\nu\mu} - \partial_\mu \phi \big ).
\eqno (A.1)
\]
Using the simple rules of factorization, we obtain the following from the above equation:
\[
{\cal L}_{\bar B} - {\cal L}_B = \frac{1}{2} \, \big (B^\mu + \bar B^\mu \big )\, \big (B_\mu - \bar B_\mu \big ) + \big (B^\mu + \bar B^\mu \big )\, 
(\partial_\mu \phi)
- \big (B^\mu - \bar B^\mu) \,\big(\partial^\nu B_{\nu\mu} \big ).
\eqno (A.2)
\]
It is straightforward to note that,  after a bit of re-arrangements, we have the following:
\begin{eqnarray*}
{\cal L}_{\bar B} - {\cal L}_B &=& \frac{1}{2} \, \big (B^\mu + \bar B^\mu \big )\, \big (B_\mu - \bar B_\mu  + 2 \, \partial_\mu \phi \big ) 
- \big (B^\mu - \bar B^\mu + 2\, \partial^\mu \phi \big ) \,\big(\partial^\nu B_{\nu\mu} \big ) \nonumber\\
&\equiv& \big [B^\mu - \bar B^\mu + 2\, \partial^\mu \phi \big ] \, \Big [\frac{1}{2} \big (B_\mu + \bar B_\mu \big ) - \partial^\nu B_{\nu\mu} \Big ].
~~~~~~~~~~~~~~~~~~~~~~~~~ ~~~~~~~~~~(A.3)
\end{eqnarray*}
In the above, it can be readily seen that we have added a total spacetime derivative term: $\partial_\mu \big [-2\,(\partial_\nu B^{\nu\mu})\,\phi \big ] $
which does {\it not} make any difference (in any manner) as far as the dynamics of our theory,
emerging out from the Lagrangian densities, is concerned.

We conclude this short Appendix with a couple of final remarks. First, the difference between the Lagrangian density [cf. Eq. (36)] and
the Lagrangian density [cf. Eq. (16)] is {\it zero} if we invoke the validity of the CF-type restriction [cf. Eq. (A.3)]. Second,
it is because of the CF-type restriction (i.e. $B_\mu - \bar B_\mu + 2\, \partial_\mu \phi = 0$) that (i)  both the
Lagrangian densities ${\cal L}_B$ and ${\cal L}_{\bar B}$ are called as the {\it coupled} Lagrangian densities [cf. Eq. (A.3)], and (ii)
both the 
Lagrangian densities ${\cal L}_B$ and ${\cal L}_{\bar B}$ 
are called as {\it equivalent} because they respect {\it both} the BRST and anti-BRST symmetry transformations [cf. Eqs. (18),(38),(46),(47)].\\

\vskip 0.3cm

\begin{center}
{\bf Appendix B: On the Anticommutativity of the (Anti-)BRST Charges}\\
\end{center}

\vskip 0.3cm

\noindent
Our present Appendix is devoted to the derivation of the CF-type restriction (i.e. $B_\mu - \bar B_\mu + 2 \,\partial_\mu \phi = 0 $) in 
the physical-sector
from the requirement of the validity of the absolute anticommutativity (i.e. $\{ Q_B, \; Q_{AB} \} = 0 $)
between the off-shell nilpotent versions of the (anti-)BRST charges $Q_{(A)B} $. In this context, we observe that
the following are {\it true}, namely;
\begin{eqnarray*}
~~~~~~~~~~~~s_b Q_{AB} &=& \int d^2 x\, \Big [ \big (\partial^0 B^i - \partial^i B^0 \big )\,\bar B_i - 
\big (\partial^0 \bar B^i - \partial^i \bar B^0 \big)\, B_i
+ 3\, \big (\dot \rho \, \lambda - \rho \, \dot \lambda \big ) \Big ], \nonumber\\
~~~~~~~~~~~s_{ab} Q_{B} &=& \int d^2 x\, \Big [ \big (\partial^0 B^i - \partial^i B^0 \big )\,\bar B_i - 
\big (\partial^0 \bar B^i - \partial^i \bar B^0 \big)\, B_i
+ 3\, \big (\dot \rho \, \lambda - \rho \, \dot \lambda \big ) \Big ],
~~(B.1)
\end{eqnarray*}
where we have directly applied (i) the BRST symmetry transformations (17) on the off-shell nilpotent expression for the anti-BRST charge in (57),
and (ii) the anti-BRST symmetry transformations (37) on the nilpotent version of the BRST charge (32).
There is a reason behind the equality: $s_b Q_{AB} = s_{ab} Q_{B} $. This is due to the fact that {\it both} the variations are equal to,
primarily, the same anticommuttaor (i.e. $\{ Q_B, \; Q_{AB} \} \equiv \{ Q_{AB}, \; Q_{B} \} $). We dwell a bit on the derivation of 
(B.1) by the {\it direct} application of (17) on the expression (57). It turns out, in this connection, that the following is true, namely; 
\begin{eqnarray*}
~~~~~~~~~~~~~~~~~~~~s_b Q_{AB} &=& \int d^2 x\, \Big [ \big (\partial^0 B^i - \partial^i B^0 \big )\,\bar B_i - 
\big (\partial^0 \bar B^i - \partial^i \bar B^0 \big)\, B_i
+ 2 \dot \lambda \, \rho - \lambda \, \dot \rho  \nonumber\\
&-& 2\, \rho \, \dot \lambda +  2\, \big (\partial^0 \bar C^i - \partial^i \bar C^0 \big )\, \partial_i \lambda 
+ \big (\partial^0 \bar C^i - \partial^i \bar C^0 \big )\, \partial_i \rho \Big ].  
~~~~~~~(B.2)
\end{eqnarray*}
Using (i) the Gauss divergence theorem in the evaluation of the {\it last} two terms,  and
(ii) using the following EL-EoMs that emerge out from ${\cal L}_{B}$ and/or ${\cal L}_{\bar B}$, namely;
\[
\dot \rho = \partial_i \big (\partial^i \bar C^0 - \partial^0 \bar C^i \big ), \qquad \dot \lambda = -\; \partial_i\, (\partial^i  C^0 - \partial^0  C^i),
\eqno (B.3)
\]
it is {\it not} very difficult, ultimately, to obtain the r.h.s. of the equation (B.1).

We have come to a point where we wish to focus on the non-ghost sector of the r.h.s. of (B.1). It is quite straightforward to note that the
following is true, namely;
\begin{eqnarray*}
&& \int d^2 x\, \Big [ \big (\partial^0 B^i - \partial^i B^0 \big )\,\bar B_i - 
\big (\partial^0 \bar B^i - \partial^i \bar B^0 \big) \, B_i  = \int d^2 x\, \Big [ \Big (\partial^0 [B^i - \bar B^i + 2 \partial^i \phi] \nonumber\\
&& - \;  \partial^i [B^0 - \bar B^0 + 2 \partial^0 \phi] \Big ) \,\bar B_i + \big (\partial^0 \bar B^i - \partial^i \bar B^0 \big )\,\bar B_i
- \big (\partial^0 \bar B^i - \partial^i \bar B^0 \big )\, B_i \Big ].
~~~~~~~~~~~~~~~~~(B.4)
\end{eqnarray*}
The above equation can be further re-arranged to yield the following:
\begin{eqnarray*}
&& \int d^2 x\, \Big [ \big (\partial^0 B^i - \partial^i B^0 \big )\,\bar B_i - 
\big (\partial^0 \bar B^i - \partial^i \bar B^0 \big)\, B_i \Big ] = \int d^2 x\, \Big [ \Big (\partial^0 [B^i - \bar B^i + 2 \partial^i \phi] \nonumber\\
&& - \;  \partial^i [B^0 - \bar B^0 + 2 \partial^0 \phi] \Big ) \,\bar B_i - \big (\partial^0 \bar B^i - \partial^i \bar B^0 \big )\,
\big ( B_i - \bar B_i + 2\, \partial_i \phi \big ) \nonumber\\
&& + \, 2\, \big (\partial^0 \bar B^i - \partial^i \bar B^0 \big )\, \partial_i \phi \Big ].
~~~~~~~~~~~~~~~~~~~~~~~~~~~~~~~~~~~~~~~~~~~~~~~~~~~~~~~~~~~~~~~~~~~~~~~~~~~(B.5)
\end{eqnarray*}
In the above equation, we point out that all the terms contain the components of the CF-type restriction: $B_\mu - \bar B_\mu + 2 \, \partial_\mu \phi = 0$
except the {\it last} term. As far as the evaluation of this last term is concerned, we can exploit the theoretical strength of the Gauss
divergence theorem and the following EL-EoM
\[
\big (\partial^0 \bar B^i - \partial^i \bar B^0 \big ) = -\, \varepsilon^{0ij} \, \partial_j {\cal B},
\eqno (B.6)
\]
to show that the {\it last} term is equal to zero. In other words, we observe:
\[
2\, \int d^2 x \, \big (\partial^0 \bar B^i - \partial^i \bar B^0 \big )\, \partial_i \phi = +\, 2\, \int d^2 x\, 
\big [\varepsilon^{0ij}\, \partial_i \partial_j {\cal B} \big ]\, \phi = 0.
\eqno (B.7)
\]
Thus, ultimately, we have obtained  the following {\it alternative}  expression for (B.1), namely;
\begin{eqnarray*}
~~~~~~~~~~~~s_b Q_{AB} &=& \int d^2 x\,  \Big [ \Big (\partial^0 [B^i - \bar B^i + 2\, \partial^i \phi] 
 - \;  \partial^i [B^0 - \bar B^0 + 2\, \partial^0 \phi] \Big )\,\bar B_i \nonumber\\
&-& \big (\partial^0 \bar B^i - \partial^i \bar B^0 \big )\,
\big ( B_i - \bar B_i + 2\, \partial_i \phi \big ) 
 + 3\, \big (\dot \rho \, \lambda - \rho \, \dot \lambda \big ) \Big ].
~~~~~~~~~~~~~~~~~~~~(B.8)
\end{eqnarray*}
We note that the {\it first} three terms contain the components of the CF-type restriction (i.e. $B_\mu - \bar B_\mu + 2 \, \partial_\mu \phi = 0$)
in the non-ghost sector 
and the last two terms  belong to the ghost-sector. The observation in (B.8) is the reflection of our observations in (44).

We wrap-up this Appendix with a clinching remark. We note that the requirement of the absolute anticommutativity 
(i.e. $\{ Q_B, \; Q_{AB} \} = 0 $)
of the off-shell nilpotent versions of the (anti-)BRST charges $Q_{(A)B} $ requires the existence of the CF-Type restriction which
is the reflection of our observation in (44) in connection with the requirement of the absolute anticommutativity (i.e. $\{ s_b, \; s_{ab} \} = 0 $)
between the BRST and anti-BRST symmetry transformations in the case of the gauge field $B_{\mu\nu} $. On the other hand, as far as our 
observations in (44) in the contexts of  the (anti-)ghost fields $(\bar C_\mu)C_\mu $ are concerned, they are reflected in the {\it last} two terms of (B.8)
with the proper factor of 3 (which is present in  the specific anticommutators: $\{ s_b, \; s_{ab} \} \, C_\mu = 3\, \partial_\mu \lambda, \; 
\{ s_b, \; s_{ab} \} \, \bar C_\mu = 3\, \partial_\mu \rho $). In other words, we have already derived the CF-type restriction
in the non-ghost sector of our theory. \\  

\vskip 0.9cm

\begin{center}
{\bf Appendix C: On the Anticommutativity of the (Anti-)co-BRST Charges}\\
\end{center}

\vskip 0.8cm

\noindent
The purpose of this Appendix is to establish that the conserved and  nilpotent (anti-)co-BRST charges [cf. Eq. (66)] anticommute with
each-other (i.e. $\{ Q_d, \; Q_{ad} \} = 0 $) modulo specific factors in the ghost-sector  
as do the off-shell nilpotent (anti)-co-BRST symmetry transformations $s_{(a)d} $ that have been listed
in the equation (69). To accomplish the {\it above}  goal, we exploit the following standard relationships between the continuous symmetry
transformations as the (anti-)co-BRST symmetry transformations [i.e. (60),(61)]  and the conserved 
 Noether (anti-)co-BRST charges (66) as their generators, namely;
\begin{eqnarray*}
~~~~~~~~~~~ s_d \,Q_{ad} = - i\, \{Q_{ad}, \; Q_d \}, \qquad \qquad s_{d} \,Q_{ad} = - i\, \{Q_{d}, \; Q_{ad} \},
~~~~~~~~~~~~~~~~~~~~~~~~(C.1)
\end{eqnarray*}
and compute the l.h.s. of the above equations by {\it directly} applying the 
off-shell nilpotent transformations in (60) and (61) on the  off-shell nilpotent (anti-)co-BRST charges
that have been quoted in equation (66). First of all, let us focus on the following
\begin{eqnarray*}
~~~~~~~~ s_d \,Q_{ad} = \int d^2 x\, \Big [ \rho\, \dot \lambda + {\cal B} \, \dot {\cal B} - (\partial^0 \bar C^i - \partial^i \bar C^0)\, \partial_i \lambda
+ \varepsilon^{0ij}\, B_i \, \partial_j B \Big ],
~~~~~~~~~~~~~~~~~~~~(C.2)
\end{eqnarray*}
which emerges out due to the application of (61) on $Q_{ad}$ [cf. Eq. (66)]. Using the celebrated Gauss divergence theorem and  the following
EL-EoMs
\begin{eqnarray*}
~~~~~~~~\dot {\cal B} = - \, \varepsilon^{0ij}\, \partial_i B_j, \quad \qquad \dot \rho = \partial_i\, (\partial^i \bar C^0 - \partial^0 \bar C^i), \quad \qquad
\dot \lambda = - \;\partial_i\, (\partial^i  C^0 - \partial^0  C^i),
~~~(C.3)
\end{eqnarray*}
it is straightforward to note that we can re-express (C.2) as follows
\begin{eqnarray*}
~~~~~~~~~~~~~~~~~~~~~~ s_d \,Q_{ad} = - i\, \{Q_{ad}, \; Q_d \} \equiv -\, \int d^2 x\, \Big [ \dot \rho\, \lambda - \rho\, \dot \lambda  \Big ].
~~~~~~~~~~~~~~~~~~~~~~~~~~~(C.4)
\end{eqnarray*}
Following the similar lines of arguments, it is {\it not} difficult to show that
\begin{eqnarray*}
~~~~~~~~~~~~~~~~~~~~~~~ s_{ad} \,Q_{d} =  - i\, \{Q_{d}, \; Q_{ad} \} \equiv -\, \int d^2 x\, \Big [ \dot \rho\, \lambda - \rho\, \dot \lambda  \Big ].
~~~~~~~~~~~~~~~~~~~~~~~~~~~(C.5)
\end{eqnarray*}
The variations $s_d \,Q_{ad} $ and $s_{ad} \,Q_{d} $ are {\it equal} because both of them correspond to the {\it same} anticommutator
(i.e. $\{Q_{ad}, \; Q_d \} \equiv \{Q_{d}, \; Q_{ad} \} $). It is interesting to point out that
 the {\it same} integral [i.e. $(C.4)$ and/or $(C.5)$], modulo a factor of $ -\,3$, appears 
[cf. Eq. (B.8)] in the ghost-sector\footnote{As a side remark, we would like to mention that if we make the choices: $C_0 = 0 $ and $\bar C_0 = 0 $
in our equations (44) and (69), we shall obtain $\dot \lambda = 0$ and  $\dot \rho = 0$. Under these conditions, we shall obtain
the absolute anticommutativity between (i) the (anti-)co-BRST charges, and (ii) the (anti-)BRST charges. However, the {\it latter} will
be satisfied if and only if we invoke the CF-type restriction in the physical-sector. It is worthwhile to point out here that the 
above choices: $C_0 = 0,\; \bar C_0 = 0$ would imply that the EL-EoMs in (B.3) and/or (C.3) would lead to $\dot \lambda = 0, \;\dot \rho = 0$
if and only if we invoke: $\partial_i C^i = 0, \; \partial_i \bar C^i = 0 $.}
of the expression for the requirement of the absolute anticommutativity between the off-shell nilpotent
versions of the BRST charge $Q_B$ and anti-BRST charge $Q_{AB}$ in our Appendix B. This is precisely due to our observations in (69) and (44).\\

\vskip 0.3cm

\begin{center}
{\bf Appendix D: On the Derivation of the Bosonic Charge $Q_w$}\\
\end{center}

\vskip 0.3cm

\noindent
The central objective of our present Appendix is to derive the {\it unique} bosonic charge $ Q_w = \{Q_B, \; Q_d \} = -\, \{Q_{AB}, \; Q_{ad} \}$
that is present in (91) which is equivalent to $Q_w = \{Q_b, \; Q_d \} = -\, \{Q_{ab}, \; Q_{ad} \} $. Toward
this goal in mind,  we exploit the theoretical strength of the relationship between the continuous symmetries and their generators as
the conserved charges [cf. Eq.(22)]. For instance, for the sake of brevity, 
we compute the l.h.s. of the relationship: $s_d Q_b = -i\, \{Q_b, \; Q_d \}$
by {\it directly} applying the co-BRST symmetry transformations  (61) on the explicit expression for the 
Noether BRST charge $Q_b$  [cf. Eq. (21)] which leads to the following explicit expression, namely;
\begin{eqnarray*}
~~~~~~~~~~~~s_d \,Q_b = \int d^2 x\, \Big [  F^{0i} \, \partial_i {\cal B} + (\partial^0 \bar C^i -  \partial^i \bar C^0 )\, \partial_i \lambda 
+ B \, \dot {\cal B} 
- \rho\, \dot \lambda \Big ].
~~~~~~~~~~~~~~~~~~~~~(D.1)
\end{eqnarray*}
A comparison between the above expression and the expression for the bosonic charge [cf. Eq. (75)] shows that 
the first {\it two} terms of both these expressions are {\it same}. In other words, we have already derived the first two terms of the 
{\it unique} bosonic charge (75). At this juncture, we concentrate on the remaining two terms of (D.1) and exploit  the
theoretical strength of the Gauss divergence theorem and appropriate EL-EoMs
in order to obtain our desired result. In this context, first of all, we focus on the following
EL-EoMs
\begin{eqnarray*}
~~~~~~~~~~~~\partial_{\mu}(\partial^{\mu}  {C}^{\nu} - \partial^{\nu}  {C}^{\mu} ) + \partial^{\nu} \lambda &=& 0 \qquad  \Rightarrow \quad
\partial_i \big (\partial^i C^0 - \partial^0 C^i \big ) = -\, \dot \lambda, \nonumber\\
\varepsilon^{\mu \nu \sigma} \,\partial_{\nu}\, {B}_{\sigma} + \partial^{\mu} {\cal {B}} &=& 0 \qquad \Rightarrow \qquad
\dot {\cal B} = -\, \varepsilon^{0ij}\, \partial_i B_j,
~~~~~~~~~~~~~~~~~~~(D.2)
\end{eqnarray*}
that emerge out from the Lagrangian density ${\cal L}_B$.
The substitution of the above into (D.1) leads to the following form of the l.h.s. of (D.1), namely;
\[
s_d \,Q_b = \int d^2 x\, \Big [ F^{0i} \, \partial_i {\cal B} + (\partial^0 \bar C^i -  \partial^i \bar C^0 )\, \partial_i \lambda 
- \varepsilon^{0ij}\, (\partial_i B_j) \, B  
+ \rho\, \partial_i \big (\partial^i C^0 - \partial^0 C^i \big ) \Big ].
\eqno (D.3)
\]
At this stage, we apply the Gauss divergence theorem to recast the above equation in the following desired form
\[
s_d \,Q_b = \int d^2 x\, \Big [ F^{0i} \, \partial_i {\cal B} + (\partial^0 \bar C^i -  \partial^i \bar C^0 )\, \partial_i \lambda 
- \varepsilon^{0ij}\,  B_i \, \partial_j B  
-   \big (\partial^0 C^i - \partial^i C^0 \big )\,\partial_i \rho  \Big ],
\eqno (D.4)
\]
where we have exploited the anticommutativity property (i.e. $\lambda \, \rho +   \rho\, \lambda = 0$)
between $\rho$ and $\lambda$. A close look at the above
expression and (75) establishes that we have already derived the {\it unique} bosonic charge $Q_w$ [cf. Eq. (75)]. We would 
like to lay emphasis on the fact that  we have started out with the relationship $s_d Q_b = -i\, \{Q_b, \; Q_d \}$ and ended
up obtaining the expression for $Q_w$ on the r.h.s. of (D.4). In other words, we have obtained 
\[
s_d \,Q_b = -i\, \{Q_b, \; Q_d \} = Q_w, 
\eqno (D.5)
\]
which demonstrate that we differ by a factor of $- \, i$ from what we have written in (91). However, this is {\it not} a serious problem
because one can get rid of {\it this} factor by taking into account an overall factor of $-\, i$ in  all the infinitesimal
variations of the fields 
that are present in the co-BRST symmetry transformations (61). This consideration  will lead to: 
$ s_d \,Q_b = -i\, \{Q_b, \; Q_d \} = -\, i\,Q_w$
in the equation (D.5). As a consequence, we  shall obtain our desired relationship: $  \{Q_b, \; Q_d \} = Q_w $ which is present in (91).

For the sake of the completeness of our discussion, we now concentrate on the computation of the l.h.s. of the relationship:
$s_{ad} Q_{ab} = -i\, \{Q_{ab}, \; Q_{ad} \}$ which will {\it also} lead to the derivation of the {\it unique} bosonic charge $Q_w$
of our theory. To accomplish this goal, we {\it directly} apply the anti-co-BRST symmetry transformations (60) on explicit
expression for the anti-BRST charge $Q_{ab}$ [cf. Eq. (41)]. This operation leads to: 
\begin{eqnarray*}
~~~~~~~~~~~s_{ad} \,Q_{ab} = -\,\int d^2 x\, \Big [  F^{0i} \, \partial_i {\cal B} + B \, \dot {\cal B} - (\partial^0  C^i -  \partial^i  C^0 )\, \partial_i \rho 
- \lambda\, \dot \rho \Big ].
~~~~~~~~~~~~~~~~~~(D.6)
\end{eqnarray*}
It is obvious that the {\it first} and {\it third} terms of the above expression match with the {\it same} terms that are
present in the explicit expression for $Q_w$ in (75). In other words, we have already derived the first and third terms of
$Q_w$ modulo a sign factor. At this stage, we focus on the remaining {\it two} terms of (D.6) and use the following
EL-EoM
\begin{eqnarray*}
~~~~~~~~~~~~\partial_{\mu}(\partial^{\mu}  \bar {C}^{\nu} - \partial^{\nu} \bar {C}^{\mu} ) - \partial^{\nu} \rho = 0 \qquad  \Rightarrow \quad
\partial_i \big (\partial^i \bar C^0 - \partial^0 \bar C^i \big ) = +\, \dot \rho, 
~~~~~~~~~~~~~~~~~(D.7)
\end{eqnarray*}
to express the last entry of (D.6) and take the help of $\dot {\cal B}$ [that is present in (D.2)] to re-write the equation (D.6) as follows:
\begin{eqnarray*}
s_{ad} \,Q_{ab} = -\,\int d^2 x\, \Big [  F^{0i} \, \partial_i {\cal B} -  \varepsilon^{0ij} \, B\, \partial_i B_j - (\partial^0  C^i -  \partial^i  C^0 )\, \partial_i \rho 
- \lambda\, \partial_i \big (\partial^i \bar C^0 - \partial^0 \bar C^i \big ) \Big ].~~(D.8)
\end{eqnarray*}
Using the Gauss divergence theorem in the {\it second} and {\it fourth} terms and taking into account the anticommutatvity property 
(i.e. $\rho \, \lambda + \lambda\, \rho = 0 $) between the fermionic auxiliary fields $ \rho$ and $\lambda$, we can re-write the
equation (D.8) in the following form
\begin{eqnarray*}
s_{ad} \,Q_{ab} = -\,\int d^2 x\, \Big [  F^{0i} \, \partial_i {\cal B} -  \varepsilon^{0ij} \, B_i\, \partial_j B - (\partial^0  C^i -  \partial^i  C^0 )\, \partial_i \rho 
+   \big (\partial^0 \bar C^i - \partial^i \bar C^0 \big )\, \partial_i \lambda \Big ],~(D.9)
\end{eqnarray*}
which, ultimately, implies: $s_{ad} \,Q_{ab} = -\, Q_w $ because the r.h.s. of the above equation matches with the expression for
$Q_w$ [cf. Eq. (75)] modulo a sign factor. In other words, we have obtained: $-i\, \{Q_{ab}, \; Q_{ad} \} = -\, Q_w $ which differs from what we have
written in (91) modulo a factor of $-\, i$. This is {\it not} a serious issue because one can get rid of this factor by incorporating 
an overall $-\, i$ factor in the anti-co-BRST symmetry transformations 
(for all the fields of our theory) that are present in (60). This will enable us to
obtain the relationship: $s_{ad} \,Q_{ab} = - i\, \{Q_{ab}, \; Q_{ad} \} = +\,i\, Q_w $.
As a consequence, we   obtain our desired  relationship: $ \{Q_{ab}, \; Q_{ad} \} =  -\, Q_w $ 
which is  written in (91).

We end this Appendix with a couple of crucial remarks. First of all, we note that operations (i.e. $s_{ad} \,Q_{AB} $ and $s_{d} \,Q_{B} $)
of the (anti-)co-BRST symmetry
transformations $s_{(a)d} $ on the off-shell nilpotent versions of the (anti-)BRST charges $Q_{(A)B}$
 {\it also} lead to the derivation of the precise expression for $Q_w$ provided we exploit the Gauss divergence theorem along with the EL-EoMs
(D.2) and (D.7).
Second, we would like to point out that the relationships: $s_{ab} \,Q_{ad} = - i\, \{Q_{ad}, \; Q_{ab} \} $ 
and $s_{b} \,Q_{d} = - i\, \{Q_{d}, \; Q_{b} \} $ also define $Q_w$ in a precise manner provided we use the EL-EoMs (D.2) and (D.7)
along with the EL-EoM: $\partial_\mu F^{\mu\nu} + \partial^\nu B = 0$ and the Gauss divergence theorem at appropriate places.\\

\end{document}